\documentclass[10pt,journal,compsoc]{IEEEtran}

\ifCLASSOPTIONcompsoc
  \usepackage[nocompress]{cite}
\else
  \usepackage{cite}
\fi

\ifCLASSINFOpdf
  \usepackage[pdftex]{graphicx}
\else
  \usepackage[dvips]{graphicx}
\fi

\usepackage{amsmath}
\usepackage{algorithmic}
\usepackage{array}

\ifCLASSOPTIONcompsoc
  \usepackage[caption=false,font=normalsize,labelfont=sf,textfont=sf]{subfig}
\else
 \usepackage[caption=false,font=footnotesize]{subfig}
\fi

\usepackage{balance}
\usepackage{color}
\usepackage{url}
\usepackage{booktabs}
\usepackage{multirow}

\begin{document}
\bstctlcite{IEEEexample:BSTcontrol}
% \title{Does Visual Self-Supervision Improve Learning of Speech Representations?}

\title{Does Visual Self-Supervision Improve Learning of Speech Representations for Emotion Recognition?}

\author{Abhinav~Shukla, Stavros~Petridis~\IEEEmembership{Member,~IEEE} and~Maja~Pantic,~\IEEEmembership{Fellow,~IEEE}% <-this % stops a space
\IEEEcompsocitemizethanks{\IEEEcompsocthanksitem A. Shukla, S. Petridis and M. Pantic are with the iBUG group in the Department of Computing at Imperial College London.\protect\\
E-mail: a.shukla@imperial.ac.uk
\IEEEcompsocthanksitem S. Petridis is also with Facebook, London, UK.
\IEEEcompsocthanksitem M. Pantic is also with Facebook, London, UK.}% <-this % stops an unwanted space
\thanks{Accepted for publication in IEEE Transactions on Affective Computing.}}

% \markboth{IEEE Transactions on Affective Computing,~Vol.~??, No.~?, October~2020}%
% {Shukla \MakeLowercase{\textit{et al.}}: Does Visual Self-Supervision Improve Learning of Speech Representations?}

\markboth{IEEE Transactions on Affective Computing, March~2021}%
{Shukla \MakeLowercase{\textit{et al.}}: Does Visual Self-Supervision Improve Learning of Speech Representations for Emotion Recognition?}


\IEEEtitleabstractindextext{%
\begin{abstract}
Self-supervised learning has attracted plenty of recent research interest. However, most works for self-supervision in speech are typically unimodal and there has been limited work that studies the interaction between audio and visual modalities for cross-modal self-supervision. This work (1) investigates visual self-supervision via face reconstruction to guide the learning of audio representations; (2) proposes an audio-only self-supervision approach for speech representation learning; (3) shows that a multi-task combination of the proposed visual and audio self-supervision is beneficial for learning richer features that are more robust in noisy conditions; (4) shows that self-supervised pretraining can outperform fully supervised training and is especially useful to prevent overfitting on smaller sized datasets. We evaluate our learned audio representations for discrete emotion recognition, continuous affect recognition and automatic speech recognition. We outperform existing self-supervised methods for all tested downstream tasks. Our results demonstrate the potential of visual self-supervision for audio feature learning and suggest that joint visual and audio self-supervision leads to more informative audio representations for speech and emotion recognition.
\end{abstract}

}

\maketitle

\IEEEraisesectionheading{\section{Introduction}\label{sec:introduction}}
\IEEEPARstart{D}{eep} neural networks trained in a supervised manner are a popular contemporary choice for various speech related tasks such as automatic speech recognition (ASR), emotion recognition and age/gender recognition. However they are a double-edged sword by virtue of providing extremely good performance given that large scale annotated data is available, which is usually expensive. For problems like emotion recognition, reliably annotated data is also extremely scarce and even modern datasets are very limited in size. Transfer learning approaches attempt to solve this problem by domain adaptation (e.g. using supervised ImageNet pretraining for downstream visual tasks), but even they need a large amount of annotated data for the primary supervised task and generalization is not guaranteed. Self-supervised learning is a recent and rapidly developing area of machine learning which might offer a potential solution to this problem.
% Recent self-supervised approaches like MoCo \cite{he2019momentum} and GDT \cite{patrick2020multi} are extremely interesting ways to learn representations from unlabeled data. However, they present results that are inferior to fully supervised pretraining. PIRL \cite{misra2019self} is an extension of MoCo that outperforms fully supervised pretraining on ImageNet for object detection.
In this work, we present a method for visually guided self-supervised learning of speech features that outperforms baseline self-supervised methods and also outperforms fully supervised pretraining on the evaluated downstream tasks.

Self-supervision is an interesting way to attempt to combat the paucity of labeled data by capturing the intrinsic structure of the unlabelled data. The idea behind self-supervision is to find a `pretext task / proxy task' for the network to learn that does not require any explicit labeling, but instead the data's inherent structure \textit{provides} the labels. During training, the network is tasked with predicting these implicit labels, which could be of various kinds. For instance, predicting the next element or a randomly masked element of a known sequence given the history/context is a popular pretext task. The key idea is that the whole sequence is already available as an unlabeled data sample, and we are just choosing an intrinsic property (here the value of the element to be predicted) as the label for the proxy supervised learning problem. This `label' is provided to us for free by the data and does not require any sort of external annotation. These pretext tasks may also model and span across multiple modalities (e.g. predicting the data or features of one modality from another). This is especially relevant in the context of affective computing where we are interested in modeling complementary multimodal information, especially in audio and video.

\begin{figure*}[t]
    \centering
    \includegraphics[width=\textwidth]{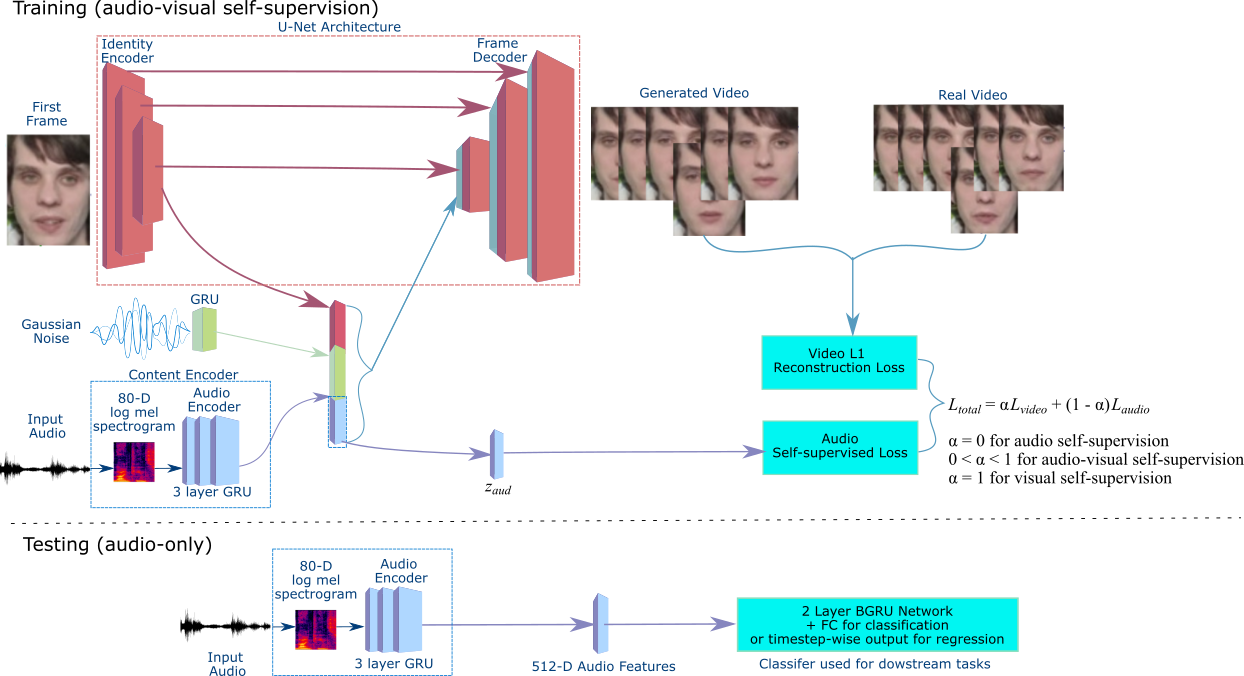}
    \caption{An overview of our proposed model for visually guided self-supervised audio representation learning. During training, we generate a video from a still face image and the corresponding audio and optimize the reconstruction loss. An optional audio self-supervised loss can be added to the total to enable multi-modal self-supervision. During testing, we use the audio encoder to extract features for (or finetune on) downstream audio-only tasks.}
    \label{fig:overview}
\end{figure*}

In this work, we investigate self-supervised learning for audio representations. Audio representations are a cornerstone of speech and affect recognition. Most affective computing applications involve the analysis of a speech signal using either handcrafted low level descriptors or through a supervised (or finetuned) neural network which directly predicts the labels of interest. However, self-supervised learning may offer better representations for these applications, especially in cases where labeled data is hard to come by and unlabeled audio data is readily available. We look into how self-supervision can be used to produce robust audio features that contain emotional information.

First, we examine the state-of-the-art in self-supervised audio feature learning which we use as baselines. We then propose a novel visual self-supervised method and a novel audio-only self-supervised method for learning audio features. We also show how visual self-supervision helps encode emotional information into the audio features.

Most existing self-supervised learning approaches are unimodal. The few existing cross-modal approaches typically have some interaction between the modalities in the latent space by pretext tasks like clustering \cite{alwassel2019self} but they do not produce an intuitive interaction between the two modalities. By contrast, our work proposes audio features that are explicitly guided by lip movements and facial expressions' reconstruction (see Fig. \ref{fig:overview}). We implicitly capture visual information related to lip movements and facial expressions in the audio features (see section \ref{sec:expression}). The visual modality is needed only during training and our audio features can be evaluated on audio-only datasets.
% Our work highlights the ability to use visual self-supervision using any audiovisual speech dataset to improve the performance of any audio-only method on a variety of problems and under various levels of noise.
\\

We summarize our research contributions as follows:
\begin{enumerate}
    \item We investigate visual self-supervision for learning audio features. We propose a novel method for visually-guided self-supervised learning of speech representations by face reconstruction (L1). The proposed speech features, which are correlated with \textbf{lip movements} and \textbf{facial expressions} due to being driven by video generation, outperform existing audio-only self-supervision approaches for speech and emotion recognition.
    \item We propose an audio-only self-supervised method (Odd One Out). This method is inspired from visual self-supervised learning and is based on temporal order verification as the pretext task. It offers competitive performance on the tested datasets.
    \item We combine the proposed audio-only and video-only supervision methods by multi-task learning. We find that the encoder trained in a multi-modal regime encodes richer information about the speech signal and yields the most effective representation that attains the best performance among all tested methods.
    \item We show that pretraining by audio-visual self-supervision produces a better weight initialization for downstream tasks than does training from scratch. This is especially relevant in the context of affective computing due to reduced overfitting and improved performance on small datasets.
    \item We show that the proposed visually-guided audio features are more robust for various levels of noise.
    % \item Our proposed visually-guided audio features offer interpretability by virtue of modeling facial and mouth movements.
\end{enumerate}

% The paper is organized as follows. Section \ref{sec:rw} studies relevant related work in self-supervised learning and multimodal learning, and we position our work with respect to existing literature in terms of its novelty and utility. Section \ref{sec:video} discusses our proposed method for visual-supervision to guide learning of speech representations by face reconstruction. Section \ref{sec:audio} introduces our proposed methods for audio-only self-supervision for speech representation learning. Section \ref{sec:data} introduces the datasets and baselines used in the work. Section \ref{sec:results} discusses the results of the various experiments we perform to validate our proposed features. Section \ref{sec:discussion} discusses the implications of our results, the limitations of our work, and possible future directions of research. Section \ref{sec:conclusion} concludes the paper.

\section{Related Work\label{sec:rw}}

To position our work with respect to existing literature and to highlight its novelty, we review prior work in: (1) Self-supervised learning, including audio, visual and cross modal methods; (2) Audiovisual speech recognition and methods that exploit both modalities for speech related tasks like emotion recognition.
% (3) We analyze the niche that our work occupies within the related work, and how our proposed features and pretraining method can be useful. 

\subsection{Self-Supervised Learning}
 Self-supervised learning is a rapidly developing field in machine learning, with the promise of being able to learn useful representations from unlabeled data. Perhaps the most seminal and widespread applications of self-supervised learning have come in natural language processing. Extremely popular recent works like ELMo \cite{peters2018deep} and BERT \cite{devlin2018bert} are based on predicting the next token of text based on the history or context. Self-supervised learning of visual features has also attracted a lot of research interest, whereas self-supervised learning of audio representations has received less attention so far. There have also been a few works on cross-modal self-supervised learning. We briefly survey these trends in the subsequent sub-sections.
 
\subsubsection{Self-supervised video feature learning}
There have been numerous recent works on visual self-supervised representation learning. Gidaris et al. \cite{gidaris2018unsupervised} predict rotations for unlabeled images that have been rotated by a known amount, which drives the features to encode information about the object shape and appearance. Other works try to predict the relative location of patches \cite{doersch2015unsupervised}, temporal order of frames in a video \cite{fernando2017self}, or use contrastive learning \cite{caron2020unsupervised, he2019momentum, misra2019self}.
% or audio-visual synchronization \cite{korbar2018cooperative, multisensory2018}.
BigBiGAN is a recent method proposed for adversarial self-supervised representation learning \cite{donahue2019large}. The work shows that more accurate and realistic reconstructions tend to produce better visual features for downstream tasks.
Cycle consistency is also a concept that has been explored for visual feature learning \cite{wang2019learning}.
DeepCluster \cite{caron2018deep} was an interesting idea which focused on clustering in the latent space based on iteratively improving labels provided by the model being trained. NoisyStudent \cite{xie2019self} was a follow up work on a similar concept, the idea being that the predictions of the model from a previous training epoch could be used as labels for the current epoch. 
S4L (Self-Supervised Semi-Supervised Learning) is another recent work which combines self-supervised learning with a small amount of labeled data to learn richer representations \cite{zhai2019s4l}.
Contrastive learning is a recent trend in self-supervised learning that is focused on separating representations of positive and negative pairs in the latent space. MoCo \cite{he2019momentum} is an important work in this area, and is based on distancing a positive pair from a large memory bank of negative examples. PIRL \cite{misra2019self} extends this idea to produce image representations that are invariant of the chosen pretext task. A more detailed overview of self-supervised methods for visual feature learning can be found in \cite{kolesnikov2019revisiting}.
We draw inspiration from visual self-supervised learning for the learning of audio features. In this work, we apply concepts from visual self-supervision to develop two audio-only self-supervised methods (see section \ref{sec:audio}) and a cross-modal self-supervised method based on visual reconstruction (see section \ref{sec:video}).

\subsubsection{Self-supervised audio feature learning}
There has also been a wave of recent work on self supervised audio-only representation learning.  CPC (Contrast Predictive Coding) \cite{oord2018representation} and APC (Autoregressive Predictive Coding) \cite{chung2019unsupervised} are similar approaches that model the next token of a speech segment given the history. Another method called LIM (Local Info Max) \cite{ravanelli2018learning} is based on maximizing the MI (mutual information) among randomly chosen windows in an unsupervised way to learn speaker embeddings. Wav2vec \cite{schneider2019wav2vec} is also an unsupervised pre-training method used in the context of speech recognition. Self supervised audio features have also been proposed for mobile devices \cite{tagliasacchi2019self}. Another very relevant recent work is PASE (Problem Agnostic Speech Encoder) \cite{pascual2019learning}, which aims to learn multi-task speech representations from raw audio by predicting a number of handcrafted features such as MFCCs, prosody and waveform.
SeCoSt \cite{kumar2019secost} is a teacher-student self-supervised approach very similar to the ones in the visual domain that iteratively use the predictions of one epoch as the labels for the next one. Phase prediction \cite{quitry2019learning} has also been proposed as an audio-based pretext task.
WaveNet \cite{chorowski2019unsupervised} is a generative model for raw audio waveforms that can be used for generic audio representations. There has also been a new version of CPC proposed for audio for multiple languages \cite{riviere2020unsupervised}. We compare our proposed methods with the best performing audio-only self-supervised baselines in recent literature. A detailed description of the baselines can be found in section \ref{sec:data} and the results can be found in section \ref{sec:results}.

\begin{figure*}[t]
    \centering
    \includegraphics[width=\textwidth]{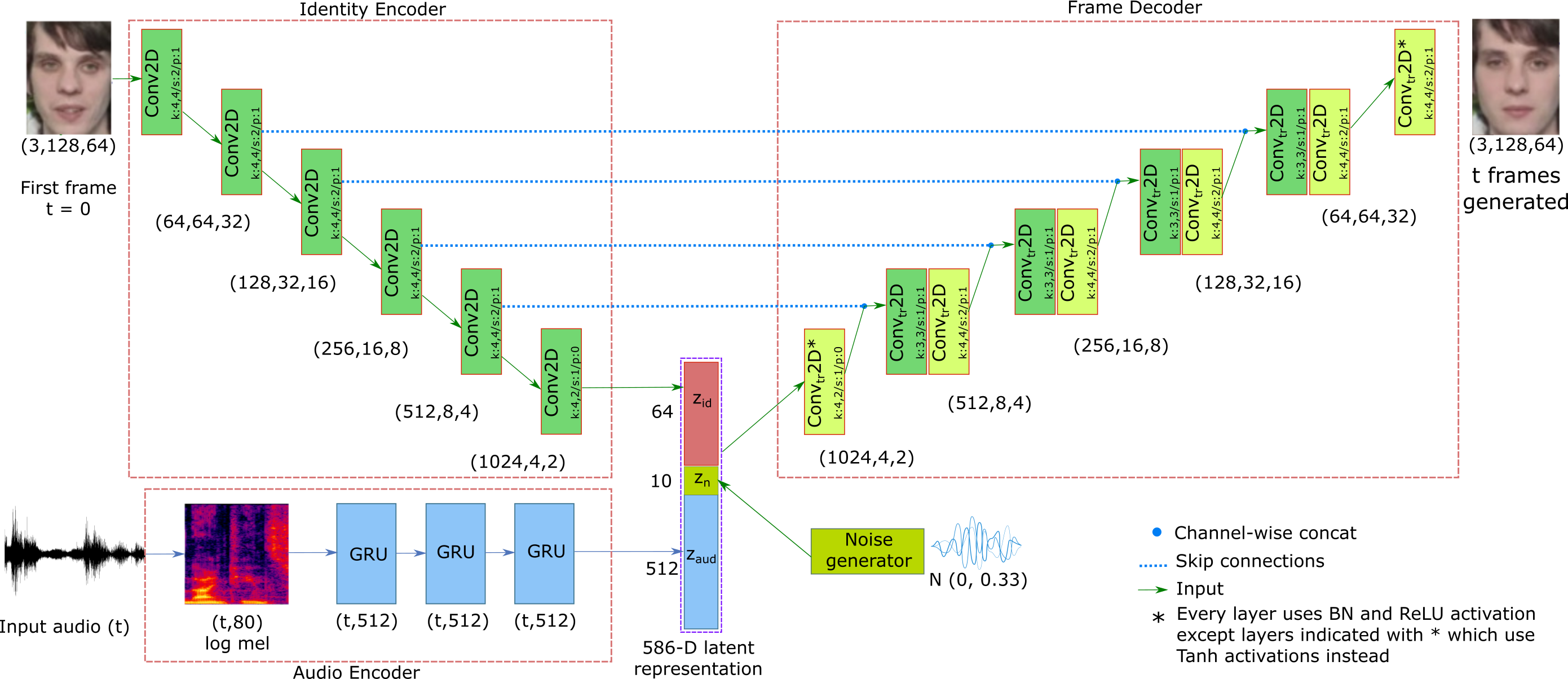}
    \caption{A detailed illustration of our the encoder-decoder model we use for video reconstruction. From an unlabeled sample of audiovisual speech, we use the audio and the first frame of the video (\textit{t = 0}) to generate a video with \textit{t} frames. The model contains: (1) an identity encoder which produces a 64-D identity embedding; (2) an audio encoder which converts the input audio (t frames of 80 dimensional log mel spectrograms) into a 512-D audio embedding; (3) a frame decoder which generates video from the concatenated latent representation using transposed convolutions.}
    \label{fig:generator}
\end{figure*}

\subsubsection{Self-supervised cross-modal learning}
A few works also exploit the relationship between modalities, such as by predicting audiovisual synchronization \cite{chung2016lip, multisensory2018}, cyclic transitions \cite{pham2019found}, the relationship between ambient sound and vision \cite{owens2018learning}, and cross-modal prediction based fusion \cite{petridis2015prediction}.
XDC \cite{alwassel2019self} extends the idea of clustering as a pretext tasks across modalities, with the cluster predictions for video coming from audio and vice versa.
Piergiovanni et al. \cite{piergiovanni2020evolving} propose a method that shares representations across modalities via distillation and finds better loss functions using evolutionary algorithms.
Contrastive Multiview Coding \cite{tian2019contrastive} is a method that projects representations of views from different modalities closer in the latent space for positive examples and further for negative examples. In this process, the encoders for each view (modality) learn useful representations.
Patrick et al. \cite{patrick2020multi} extend the contrastive learning concept from MoCo \cite{he2019momentum} to a multi-modal setting.
Morgado et al. \cite{morgado2020audio} combine audio-visual instance discrimination with cross-modal agreement for self-supervised learning.
Zhu et al. \cite{zhu2020deep} present a detailed survey of deep audio-visual learning including cross-modal self-supervised learning.
All of these works have shown that it is possible to learn robust multi-task representations from a large amount of unlabeled data that is inexpensive to obtain. We propose a novel self-supervised method based on cross-modal reconstruction to learn audio features. Our method is based on speech-driven facial reconstruction and is explained in detail in section \ref{sec:video}. Speech based reconstruction has a long history in computer vision, and has been demonstrated by works like Video Rewrite (1997) \cite{bregler1997video}, Mary 101 (2002) \cite{ezzat2002trainable} and more recently Synthesizing Obama (2017) \cite{suwajanakorn2017synthesizing}. Our contribution is to utilize speech-driven facial animation as a pretext task for self-supervised learning of audio features, which has not been explored before. We exploit the correlation between facial expressions and emotion, and that between lip movements and speech content. This leads to audio features that contain additional information about emotion and speech due to using the visual modality as a supervisory signal.

\subsection{Audiovisual Speech and Emotion Recognition}
Audiovisual speech data is extremely common and the usage of complementary information from both modalities is a popular concept in many fields of research. The McGurk effect \cite{mcgurk1976hearing} was the classic example that demonstrated the audio-visual nature of human perception of speech. The visual modality contains information that offers robustness in circumstances where the audio modality may be corrupted with noise \cite{petridis2018end}.
% Visual-only speech recognition (lipreading) is also a popular contemporary field which has seen rapid advances recently \cite{petridis2018end, chung2017lip}.

Audiovisual emotion recognition has also seen a significant amount of recent research efforts. Automatic affect recognition has a variety of applications in various fields; from detecting depression \cite{cohn2009detecting}, to more emotionally relevant advertising \cite{shukla2017evaluating, shukla2018looking, shukla2020recognition}. A lot of contemporary affect analysis approaches are based on deep neural networks that study both the visual and audio modalities \cite{kollias2019deep, zadeh2018multimodal, shukla2017affect}. However a big problem in emotion recognition is the lack of reliably annotated data for large datasets, which we try to address (implicitly) in this paper.

% \subsection{Analysis of related work}
% The current state of the art for both affect and speech recognition is largely dominated by fully supervised models which require large amounts of labeled training data. However extensive annotation is difficult and expensive, especially for problems like emotion recognition. Self-supervised learning may offer a practical solution to this issue by virtue of being able to learn from large amounts of unlabeled data.

% Most existing self-supervised learning approaches are either unimodal (for audio and video). The few cross-modal approaches typically have some interaction between the modalities in the latent space by pretext tasks like clustering, however they don't produce an interpretable and intuitive interaction between the two modalities. Our work proposes audio features that are explicitly guided by the generation of lip movements and facial expressions by reconstruction. Our proposed method is novel in its methodology and occupies an interesting niche in audio representation learning. We capture explainable visual information in the audio features, however we can evaluate our model on audio-only datasets without requiring the visual modality. Our work highlights the ability to use visual self-supervision using any audiovisual speech dataset to improve the performance of any audio-only method on a variety of problems and under various levels of noise.

\section{Visual Self-Supervision for Speech Representation Learning\label{sec:video}}

% \begin{figure}[t]
%     \centering
%     % \includegraphics[width=\columnwidth]{figures/encdec.png} \\

%     % \vspace{0.2cm}

%     % \includegraphics[width=\columnwidth]{figures/audioencoder.png}
    
%     % \vspace{0.3cm}
    
%     \includegraphics[width=\columnwidth]{figures/encoder-mel.png}
%     \caption{The architecture of the log mel spectrogram encoder. The encoder is the component which we extract features from after self-supervised pretraining.}
%     \label{fig:generator}
% \end{figure}

The proposed method is illustrated in Fig. \ref{fig:overview} and is based on our prior work on visually guided speech representation learning through speech-driven facial animation \cite{vougioukas2018end, shukla2020visually, shukla2020cvprw, shukla2020icmlw}. The model is a temporal encoder-decoder which takes a still image of a face (frame from a 25 fps video) and an audio signal as inputs and generates video frames from these. The model itself can be conceptually divided into three subnetworks (see Fig. \ref{fig:overview} and Fig. \ref{fig:generator}), namely the content/audio encoder (3 layer GRU), the identity encoder (6 layer 2D CNN) and the frame decoder (U-Net \cite{ronneberger2015u} architecture with skip connections from the identity encoder).

The architecture of the content encoder is a 3 layer GRU with log mel spectrograms as input (closely following \cite{chung2019unsupervised}), as shown in Fig. \ref{fig:generator}. Its purpose is to convert the input audio into a latent space audio feature vector $z_{aud}$. Similarly, the identity encoder (see Fig. \ref{fig:generator} top-left), which is made of 6 (Conv2D - BatchNorm - ReLU) blocks, reduces
a 64x128 input image (which is the first video frame of the audiovisual speech segment) to a 64x1 feature vector $z_{id}$.

We also use a noise generator (see Fig. \ref{fig:overview}) capable of producing noise that is temporally coherent. A 10 dimensional vector is sampled from a Gaussian distribution with mean 0 and variance of 0.33 and passed through a single-layer GRU to produce the noise sequence. This latent representation $z_n$ accounts for randomness in the face synthesis process (such as the generation of random sequential behaviour like blinks \cite{vougioukas2019realistic}), which leads to a more realistic facial reconstruction.

The latent representation is the concatenation of $ z_{aud}, z_{id}$ and $z_n $ (as shown in Fig. \ref{fig:generator}). This results in a 586 dimensional embedding. This embedding then goes through the frame decoder (see Fig. \ref{fig:generator} top-right), which is a CNN that uses strided transposed convolutions to produce the video frames. The skip connections to the identity encoder help in preserving subject identity.

An L1 reconstruction loss between a random frame from the generated video and the corresponding frame from the real video is used to train the network. The L1 loss on the pixel level is commonly used in facial reconstruction as opposed to the L2 loss which typically produces blurrier reconstructions. We use the Adam optimizer with a learning rate of 0.06 that is decayed by a factor of 0.98 every 10 epochs. Essentially, our model aims to predict the video modality (full face reconstruction) given only the audio modality and speaker identity information from the first frame. In this process, the audio encoder is driven to produce useful \textbf{speech features that correlate with mouth and facial movements} (because we need to generate these lip and facial movements using only the audio information, so the features $ z_{aud} $ must encode this in order to reduce the L1 loss). After this process of visually guided self-supervised pretraining, we simply use the trained audio encoder as a pretrained model for audio-only downstream tasks. The features extracted from this model are especially interesting to evaluate on tasks like speech recognition and emotion recognition. This is because these features are explicitly trained (guided by the visual modality) to contain information related to lip movements (highly correlated with speech content) and facial expressions (highly correlated with emotion).
% We evaluate our trained model in detail in Section \ref{sec:results}.

\begin{figure}[t]
    \centering
    \includegraphics[width=\columnwidth]{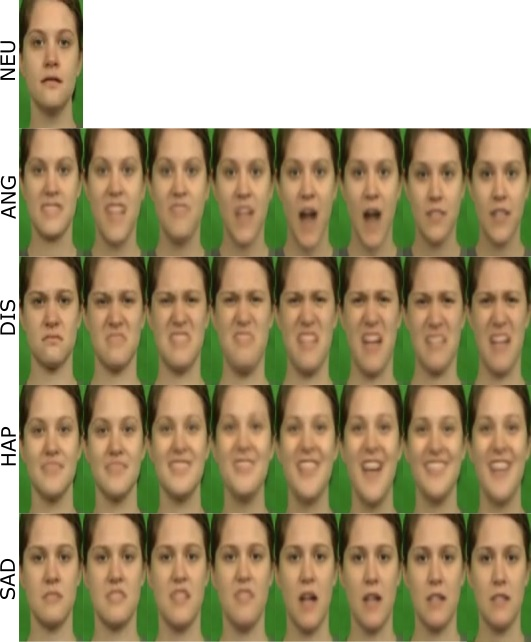}
    \caption{Videos produced by the proposed method using the same neutral still frame taken from the CREMA-D test set and driven by the sentence “its eleven o’clock” spoken with multiple emotions (Anger, Disgust, Happiness, Sadness). This highlights that emotional information is captured in the audio features being used for the generation.}
    \label{fig:emotion_generation}
\end{figure}

% \begin{figure}[t]
%     \centering
%     \subfloat[\scriptsize{\textit{Speech2Vid} \cite{chung2017you}}]{%
%       \includegraphics[width=\columnwidth]{figures/speech2face-compare.jpg}}\\
%     \subfloat[\scriptsize{Our proposed method}]{%
%       \includegraphics[width=\columnwidth]{figures/ours-compare.jpg}}
%     \caption{Qualitative comparison of the generated expressions between our method and Speech2Vid \cite{chung2017you}. Our method is better at generating expressions, due to emotional information being captured in the audio features.}
%     \label{fig:generation_comparison}
% \end{figure}

\begin{figure*}[ht]
    \centering
    \includegraphics[width=\textwidth]{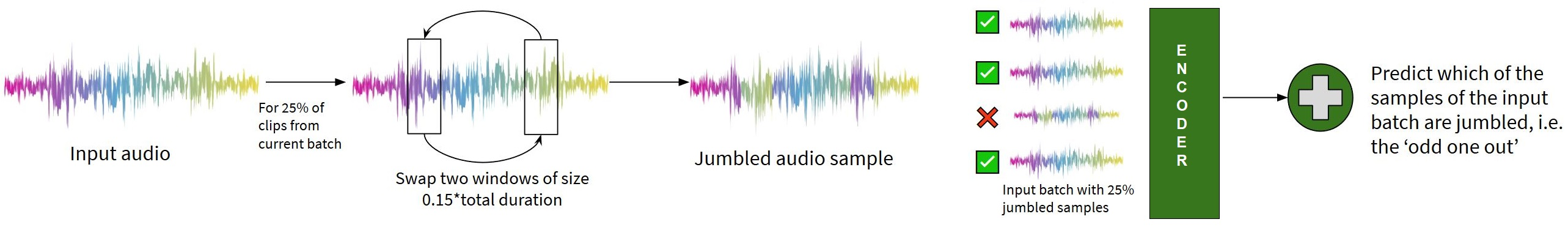}
    \caption{An overview of the proposed Odd One Out networks for audio representation learning. The input audio is jumbled as shown for 25\% of the input batch. The audio encoder is then trained on the self supervised task of predicting which clip is the `odd one out'.}
    \label{fig:oddone}
\end{figure*}

\subsection{Why does visual self-supervision produce good features for emotion?\label{sec:expression}}
To support our claim that the proposed method produces good features for emotion, we present qualitative results for full face reconstruction using our method. We use the same neutral expression still frame from the CREMA-D dataset as the input image and produce video sequences driven by different audio clips of  multiple emotions (see Fig. \ref{fig:emotion_generation}). We observe that facial expressions that correlate with the emotion of the audio clip are generated in the output video. Note that this is achieved using only a neutral still image and the audio information. This offers evidence that emotion related information is being encoded in the audio features being used to drive the generation. 
Our method is able to generate a variety of facial expressions that correlate with the emotion of the input audio. These observations form the basis of our hypothesis that the proposed task (reconstructing the full face based only on the audio information) can help produce better audio features for emotion recognition in a self-supervised manner.
The reader can find more detailed insights into the speech-driven facial animation process in our prior work \cite{vougioukas2018end, vougioukas2019realistic}.

% We also qualitatively compare the generated videos produced by our method and those by Speech2Vid \cite{chung2017you}, which is another method for speech-based face video generation (see Fig. \ref{fig:generation_comparison}). We observe that Speech2Vid is unable to generate facial expressions even for emotional audio inputs. On the other hand, our method is able to generate a variety of facial expressions that correlate with the emotion of the input audio. These observations form the basis of our hypothesis that the proposed task (reconstructing the full face based only on the audio information) can help produce better audio features for emotion recognition in a self-supervised manner.

\subsection{Audio Encoder Architecture\label{sec:encoders}}
The audio encoder (see Fig. \ref{fig:generator} bottom-left) is a log mel spectrogram encoder (closely following \cite{chung2019unsupervised}). The log mel spectrogram is computed with 80 frequency bins, a window width of 25ms and a stride of 10ms, which is a standard choice for processing speech signals. This (t, 80) dimensional input then goes through the encoder which is a 3 layer GRU network with each layer having a hidden size of 512 followed by a fully connected layer which converts it into a feature with dimensionality (t, 512). This specific architecture of the audio encoder with 3 GRU layers is the exact same as used in \cite{chung2019unsupervised}. We chose this for simplicity and to enable direct comparison with this baseline which uses a similar audio input as us (80 dimensional log mel features).

We use the above described architecture (see Fig. \ref{fig:generator}) as the audio encoder in the proposed models in the following sections (for both visual and audio self-supervision). We then use the trained encoder to extract features from the evaluation datasets. We implement our method and all evaluation pipelines using PyTorch \cite{pytorch}.

\section{Audio Self-Supervision for Speech Representation Learning\label{sec:audio}}
This section introduces an audio-only self-supervised method that we propose for speech representation learning. The concept behind it is temporal order verification for audio as a pretext task. The method is inspired by a similar work in visual self-supervision called 'Odd One Out' \cite{fernando2017self}.

% \subsection{Audio feature learning with the Arrow of Time}
% The temporal order of a sequence carries a lot of potentially useful information about its structure. For video, Wei et. al. \cite{wei2018learning} proposed the Arrow of Time as a self supervised method that predicts whether a given video sequence is being played forwards or backwards. This helps the encoder that is predicting this pretext task label to learn features that correspond to object semantics and other visually correlated physical characteristics like gravity, forces etc. which may be useful for generic visual feature learning. We adapt the Arrow of Time (henceforth abbreviated as AoT) method for speech signals. The problem reduces to predicting whether a given audio clip is being played forwards or backwards. While learning to predict the task, the encoder learns useful audio features that differentiate between certain phonemes and how they sound when played forward vs backward. In order to predict the direction of the Arrow of Time, the encoder must capture useful characteristics about the phonemes themselves. In our implementation, we simply flip the temporal order of half of the sequences of an input batch (make them play backwards), and train the audio encoder with the supervised task of predicting the binary class problem (forward or backward). We use the encoder architecture described in Section \ref{sec:encoders}. 
% The experimental setup and results are discussed in Section \ref{sec:results}.

\subsection{Odd One Out networks for Audio}
The temporal order of a sequence carries a lot of potentially useful information about its structure. Odd One Out networks for video \cite{fernando2017self} are based on predicting which one out of multiple sets of ordered sequences of frames is in jumbled order (temporally incorrect order). This helps the encoder that is predicting this pretext task label to learn features that correspond to object semantics and other visually correlated physical characteristics like gravity, forces etc. which may be useful for generic visual feature learning. We adapt this idea to the audio modality in a straightforward way. For a given input batch of audio clips, we jumble 25\% of the clips. The jumbling is performed by selecting at random two windows of a length of 15\% of the total audio duration and swapping them. The encoder is then tasked with predicting which element in the input batch is the 'Odd One Out', and is optimized using cross entropy loss. While learning to predict the task, the encoder learns useful audio features that differentiate between certain phonemes and how they sound when played jumbled vs in-order. In order to predict whether the sequence is jumbled, the encoder must capture useful characteristics about the phonemes themselves. Fig. \ref{fig:oddone} illustrates the training procedure for Odd One Out networks for audio representation learning. We use the encoder architecture described in Section \ref{sec:encoders}.
% Our results are described in Section \ref{sec:results}

\section{Audio-visual Self-supervision for Speech Representation Learning}
We combine the proposed audio and visual self-supervision methods by making the encoder jointly predict the visual self-supervision task and the audio self-supervision task. Since we used the same encoder architecture for both the visual and audio tasks, this is straightforward to accomplish. In the pipeline shown in Fig. \ref{fig:overview} for visual self-supervision, we also use the optional prediction for the audio-only self-supervised task (Odd). This leads to two losses being calculated, one for visual and one for audio self-supervision.

The total loss $ L_{total} $ is the weighted sum of the L1 reconstruction loss from visual self-supervision $ L_{video} $ and the cross entropy loss from the audio-only self supervision $ L_{audio} $. $ \alpha $ is the weight factor which controls how much of the loss term comes from which type of supervision. The total loss is given by the equation:

% \[
%  L_{total} = \alpha L_{video} + (1 - \alpha) L_{audio} 
% \]

\begin{equation}\label{eq:loss}
 L_{total} = \alpha L_{video} + (1 - \alpha) L_{audio}
\end{equation}

% We have two possible multimodal self-supervised models being trained depending audio self-supervision type, namely: L1 + AoT and L1 + Odd.

\section{Datasets and Baselines\label{sec:data}}

\subsection{Datasets}
This section introduces the various audio-only and audiovisual datasets that were used in the work either for pretraining or evaluating the baseline and proposed models. For all datasets, we divide the data into training, validation and test sets with all samples from each speaker belonging to a particular set only. Table \ref{tab:Subjects} summarizes the statistics for all the datasets used in this work.

 The CREMA-D dataset \cite{cao2014crema} contains a diverse set of 91 actors who utter 12 sentences multiples times each with a different level of intensity for each of 6 basic emotional labels (anger, fear, disgust, neutral, happy, sad). We use Crema as a discrete emotion recognition evaluation dataset.
 
 The Ravdess dataset \cite{livingstone2018ryerson} contains 1440 samples of 24 different actors who acted out two sentences with 8 different basic emotions (anger, calm, sad, neutral, happy, disgusted, surprised, fear) and two different intensity levels. We use Ravdess also as a discrete emotion recognition evaluation dataset.
 
 The RECOLA dataset \cite{ringeval2013introducing} contains dyadic conversations in French between a pair of participants working to solve a collaborative task over video conference. The annotated part of the dataset consists of 5-minute long clips (with audio from one speaker only) with continuous valence and arousal annotations from 6 annotators. Since the labels of the test set are not publicly available, we report the performance of our models on the development set.
 
 The SEWA dataset \cite{kossaifi2019sewa} contains dyadic conversations over video conference between a pair of participants discussing about an advertisement that they have just watched. The audio clips are typically 3-minute long and have continuous valence and arousal annotations. We use the AVEC2017 challenge German language subset (with the gold standard valence and arousal labels) of the dataset to run our experiments.
 
 The IEMOCAP dataset \cite{busso2008iemocap} contains dyadic conversations between 10 speakers for a total of 12 hours of audiovisual data. The discrete emotion labels comprise of 8 categories (anger, happiness, sadness, neutral, excitement, frustration, fear, surprise), however we only consider the first 4 categories for our experiments (anger, happiness, sadness, neutral). This is due to much higher inter annotator agreement for these categories, and this portion of the dataset has been similarly used in prior studies \cite{majumder2018multimodal}. This partition also leaves us with around 6.5 hours of data instead of the original 12 hours. We use IEMOCAP as another discrete emotion recognition evaluation dataset.
 
 The SPC (Speech Commands) dataset \cite{warden2018speech} contains 64,727 total utterances of 30 different words by 1,881 speakers. We use SPC as a speech recognition evaluation dataset.
 
%  The GRID dataset \cite{cooke2006audio} contains audio-visual speech recordings of subjects in full frontal view. It has 33 speakers, each of whom speak 1000 sentences containing six words. Every sentence in the GRID dataset follows a particular format for every word: [command/colour/preposition/letter/digit/adverb]. An example sentence is ``Bin blue at F 1 now". We use GRID as an ASR evaluation dataset, and use only the audio modality for WER (word error rate) evaluation.
 
 The LRW dataset \cite{chung2016lip} is a large, in-the-wild dataset of 500 different isolated words primarily from BBC recordings. It is an audiovisual speech dataset and is thus appropriate for training our methods. We use a subset of LRW that has only nearly frontal videos (with yaw, pitch and roll restricted to a maximum of 10 degrees), in order to have a cleaner supervisory signal from the visual modality. This filtering leaves us with a total of around 40 hours of usable data. We use LRW as the self-supervised pretraining dataset for all baseline and proposed methods.

\subsection{Baselines}

Since the aim of our work is to yield self-supervised audio features, we compare against other baselines focusing on the same goal. The three self-supervised methods we compare against are CPC \cite{oord2018representation}, APC \cite{chung2019unsupervised} and PASE \cite{pascual2019learning}. In our prior work \cite{shukla2020visually}, we have also used other audio-visual self-supervised methods as baselines \cite{arandjelovic2018objects, korbar2018cooperative}, but due to them not obtaining competitive performance on the target datasets (CREMA/SPC), we exclude them from this paper.

% % \subsubsection{AVENet: Single Image and Audio Correspondence}
% AVENet \cite{arandjelovic2018objects} is a two-stream audio-visual correspondence based network. One second of audio along with the middle frame of the one second segment are passed as input to the parallel streams, with a positive pair coming from the correct point in the video and a negative pair coming from a different video. The optimization is done with a contrastive loss. We use the audio stream of the network for feature extraction. Korbar et. al. \cite{korbar2018cooperative} propose an audio-visual temporal synchronization network (Cooperative) which is also a two-stream audiovisual network but has 1 second of video frames as input as opposed to a single frame in AVENet. A positive pair of audio and video samples is one that is in sync, and there are various types of out of sync negative examples in progressive order of difficulty which are optimized with a curriculum learning strategy (easy first, hard later).

% \subsubsection{Contrast Predictive Coding}
Contrast Predictive Coding (CPC) \cite{oord2018representation} is a technique that tries to model a density ratio to maximize mutual information (MI) between the target signal (random raw audio window) and the context (current raw audio window). By maximizing the MI, the method can extract the underlying latent variables that the two different parts of the signal have in common. This essentially means that the representations of related (positive) examples lie closer in the latent space, and those of negative examples lie further away.
% Wav2vec \cite{schneider2019wav2vec} refines the idea from CPC specifically for speech.

Autoregressive Predictive Coding (APC) \cite{chung2019unsupervised} is similar to CPC, however the key difference is that APC directly tries to predict the immediate future part of the signal based on the history whereas CPC tries to maximize mutual information between the target (future) and the context (present). The input features for APC are 80 dimensional log mel spectrograms with a window size of 25 ms and a step size of 10 ms. The model tries to predict the log mel spectrograms for the future windows given the history.

PASE \cite{pascual2019learning} is a raw audio encoder trained in a self supervised way to predict various different handcrafted features such as MFCC, prosody, waveform etc. While predicting these multiple tasks, the encoder learns a very robust and multi-task representation for raw audio that these tasks exemplify (e.g. prosody for emotion).

We also compare our methods against 39 dimensional MFCCs (13 coefficients, 13 deltas, and 13 delta-deltas) which act as baseline features used for supervised learning for audio. We also compare with 80 dimensional log mel spectrograms. Additionally, we also use the OpenSmile \cite{eyben2013recent} toolkit to extract LLD's (low level descriptors) based on the IS13-ComParE feature set. These features are well established baselines for audio based affective computing tasks.

\begin{table}[t]
\begin{center}
\small
\begin{tabular}{lrrrrr}
\hline
Dataset & Train & Val &  Test \\
\hline
% GRID   & 31639 / 26.4 & 6999 / 5.80 & 9976 / 8.31 \\
LRW & 112658 / 36.3 & 5870 / 1.90& 5980 / 1.90\\
SPC & 51094 / 14.2 & 6798 / 1.88 & 6835 / 1.89 \\
CREMA-D & 11594 / 9.70 & 819 / 0.70 & 820 / 0.68 \\
Ravdess & 1509 / 1.76 & 415 / 0.48 & 519 / 0.60 \\
IEMOCAP & 3548 / 4.28 &  793 / 0.95 &  942 / 1.31 \\
RECOLA & 180 / 1.00 & 150 / 0.80 & -\\
SEWA & 706 / 0.78 & 287 / 0.31 & 352 / 0.39\\
\hline
\end{tabular}
\end{center}
\caption{The number of samples and duration (number / time in hours) of speech data in the training, validation and test sets of each dataset.}
\label{tab:Subjects}
\end{table}
% 5479 total, 6.55 hours of data

\section{Experiments and Results\label{sec:results}}

\begin{table*}\centering
\small
\renewcommand{\arraystretch}{1.3}
\begin{tabular}{cccccccc}
\toprule
\multicolumn{2}{c}{\textbf{Results}} & \multicolumn{3}{c}{\textbf{Discrete Emotion}} & \textbf{Speech Recognition}\\
\cmidrule(lr){1-2} \cmidrule(lr){3-5} \cmidrule(lr){6-6}
\multicolumn{2}{c}{Pretraining Dataset} & LRW & LRW & LRW & LRW\\
\multicolumn{2}{c}{Evaluation Dataset} & \textbf{CREMA-D} & \textbf{Ravdess} & \textbf{IEMOCAP} & \textbf{SPC}\\
\cmidrule(lr){3-3} \cmidrule(lr){4-4} \cmidrule(lr){5-5} \cmidrule(lr){6-6}
% \multicolumn{3}{c}{Classifier for \textit{(t, dim)} features} & LSTM & LSTM &  LSTM &  ESPNet &  LSTM \\
\multicolumn{2}{c}{Labels} & 6 emotions & 8 emotions & 4 emotions & 30 words\\
\midrule
Method & Pretext task & F1 Score ($\uparrow$) & F1 Score ($\uparrow$) & F1 Score ($\uparrow$) & F1 Score ($\uparrow$)\\
\midrule
MFCC & None & 0.447 $\pm$ 0.012 & 0.390 $\pm$ 0.013 & 0.535 $\pm$ 0.006 & 0.912 $\pm$ 0.002\\
Log Mel & None & 0.463 $\pm$ 0.013 & 0.480 $\pm$ 0.010 & 0.572 $\pm$ 0.011 & 0.927 $\pm$ 0.003\\
LLD \cite{eyben2013recent} & None & 0.409 $\pm$ 0.015 & 0.398 $\pm$ 0.013 & 0.497 $\pm$ 0.013 & 0.643 $\pm$ 0.014 \\
CPC \cite{oord2018representation} & Audio & 0.422 $\pm$ 0.016 & 0.408 $\pm$ 0.011 & 0.517 $\pm$ 0.011 & 0.814 $\pm$ 0.019\\
PASE \cite{pascual2019learning} & Audio  & 0.488 $\pm$ 0.010 & 0.472 $\pm$ 0.010 & 0.557 $\pm$ 0.017 & 0.941 $\pm$ 0.002\\
APC \cite{chung2019unsupervised} & Audio & 0.480 $\pm$ 0.012 & 0.497 $\pm$ 0.011 & 0.592 $\pm$ 0.019 & 0.933 $\pm$ 0.003\\
\midrule
% AoT & Audio & - & - & - & - & - & - & -\\
Odd & Audio & 0.542 $\pm$ 0.016 & 0.501 $\pm$ 0.008 & 0.615 $\pm$ 0.010  & 0.919 $\pm$ 0.002\\
\midrule
L1 & Visual & 0.567 $\pm$ 0.014 & 0.560 $\pm$ 0.010 & 0.618 $\pm$ 0.009 & 0.935 $\pm$ 0.002\\
\midrule
% L1 + AoT & Audio+Visual & - & - & - & - & - & - & -\\
L1 + Odd & Audiovisual & \textbf{0.573 $\pm$ 0.012} & \textbf{0.565 $\pm$ 0.008} & \textbf{0.631 $\pm$ 0.007} & \textbf{0.944 $\pm$ 0.002}\\
\midrule
L1 + Odd (finetuned) & Audiovisual & \textbf{0.592 $\pm$ 0.010} & \textbf{0.645 $\pm$ 0.016} & \textbf{0.642 $\pm$ 0.011} & \textbf{0.953 $\pm$ 0.002}\\
\midrule
Supervised & None & 0.540 $\pm$ 0.013 & 0.528 $\pm$ 0.010 & 0.623 $\pm$ 0.010 & 0.921 $\pm$ 0.005\\

\bottomrule
\end{tabular}

% Results with Total CCC
\begin{tabular}{cccccccc}
\toprule
\multicolumn{2}{c}{\textbf{Results (continued)}} &  \multicolumn{4}{c}{\textbf{Continuous Affect Recognition}}\\
\cmidrule(lr){1-2} \cmidrule(lr){3-6}
\multicolumn{2}{c}{Pretraining Dataset} & \multicolumn{2}{c}{LRW} & \multicolumn{2}{c}{LRW}\\
\multicolumn{2}{c}{Evaluation Dataset} & \multicolumn{2}{c}{\textbf{SEWA}} & \multicolumn{2}{c}{\textbf{RECOLA}}\\
\cmidrule(lr){3-4} \cmidrule(lr){5-6}
% \multicolumn{3}{c}{Classifier for \textit{(t, dim)} features} & LSTM & LSTM &  LSTM &  ESPNet &  LSTM \\
\multicolumn{2}{c}{Labels} & Valence & Arousal & Valence & Arousal\\
\midrule
Method & Pretext task & CCC ($\uparrow$) & CCC ($\uparrow$) & CCC ($\uparrow$) & CCC ($\uparrow$)  \\
\midrule
MFCC & None & 0.349 $\pm$ 0.009 & 0.359 $\pm$ 0.010 & 0.413 $\pm$ 0.008 & 0.714 $\pm$ 0.011 \\
Log Mel & None & 0.307 $\pm$ 0.011 & 0.316 $\pm$ 0.007 & 0.396 $\pm$ 0.012 & 0.696 $\pm$ 0.011 \\
LLD \cite{eyben2013recent} & None & 0.245 $\pm$ 0.011 & 0.236 $\pm$ 0.010 & 0.271 $\pm$ 0.015 & 0.624 $\pm$ 0.010\\
CPC \cite{oord2018representation} & Audio & 0.287 $\pm$ 0.008  & 0.295 $\pm$ 0.009 & 0.320 $\pm$ 0.011 & 0.623 $\pm$ 0.012\\
PASE \cite{pascual2019learning} & Audio & 0.354 $\pm$ 0.008 & 0.343 $\pm$ 0.010  & 0.390 $\pm$ 0.008 & 0.719 $\pm$ 0.007 \\
APC \cite{chung2019unsupervised} & Audio & 0.317 $\pm$ 0.009  & 0.321 $\pm$ 0.014 & 0.383 $\pm$ 0.010 & 0.703 $\pm$ 0.008\\
\midrule
% AoT & Audio & - & - & - & - & - & - & -\\
Odd & Audio & 0.342 $\pm$ 0.008 & 0.339 $\pm$0.010 & 0.436 $\pm$ 0.012 & 0.720 $\pm$ 0.008\\
\midrule
L1 & Visual & 0.361 $\pm$ 0.008 & 0.368 $\pm$ 0.009 & 0.438 $\pm$ 0.011 & 0.733 $\pm$ 0.012\\
\midrule
% L1 + AoT & Audio+Visual & - & - & - & - & - & - & -\\
L1 + Odd & Audiovisual & \textbf{0.373 $\pm$ 0.007} & \textbf{0.371 $\pm$ 0.009} & \textbf{0.448 $\pm$ 0.009} & \textbf{0.745 $\pm$ 0.008} \\
\midrule
L1 + Odd (finetuned) & Audiovisual & \textbf{0.380 $\pm$ 0.010} & \textbf{0.383 $\pm$ 0.009} & \textbf{0.452 $\pm$ 0.012} & \textbf{0.764 $\pm$ 0.006}\\
\midrule
Supervised  & None & 0.297 $\pm$ 0.019 & 0.330 $\pm$ 0.009 & 0.402 $\pm$ 0.010 & 0.701 $\pm$ 0.008\\
\bottomrule
\end{tabular}

\caption{Results for all baseline and proposed methods for discrete emotion recognition (on CREMA, Ravdess and IEMOCAP), speech recognition (on SPC) and continous affect recognition (on SEWA and RECOLA). Unless specified otherwise, all methods are used as frozen audio feature extractors before training a classifier (BGRU) on the downstream task. All results are in the form of $\mu \pm \sigma$ for 10 runs.}
    \label{tab:all_results}
\end{table*}

This section presents the details of all experiments that we perform to rigorously validate our proposed method. We present all results for discrete emotion recognition, continuous affect recognition, and speech recognition from the extracted features in Table \ref{tab:all_results}, for both visual and audio self-supervision for all variants of the models. We also show the results with the combination of the visual and audio self-supervision approaches using multi-task learning. We present numerous ablation studies such as the variation of model performance with change in pretraining set size and noise level. We also compare the frozen encoders (that provide fixed extracted features) with their finetuned (pretrained weights are also updated in downstream training) and fully supervised (trained from scratch on the downstream task) equivalents.
% We implement all our proposed methods and evaluation pipelines using PyTorch\cite{pytorch}

\subsection{Experimental Setup}
We evaluate all extracted features on: (i) Discrete Emotion Recognition, (ii) Continuous Affect Recognition, and (iii) Automatic Speech Recognition (ASR).

For the \textbf{emotion recognition} task, we first perform self-supervised pretraining on LRW as described, and then use the pretrained models as feature extractors on the CREMA, Ravdess and IEMOCAP datasets. Once we have these features, we then train an BGRU (bidirectional gated recurrent unit) model for the emotion classification task. We used a BGRU for simplicity, however this can be replaced by any model that can classify variable length sequences into discrete categories (such as LSTMs, TCNs, Transformers, LiGRUs \cite{ravanelli2018light}). We use a 2 layer BGRU with 256 units in each layer. The initial learning rate is 0.0001 and is decayed by a factor of 0.1 every 40 epochs. We train for 100 epochs and use the checkpoint from the epoch which gives the best validation set metric for evaluation on the test set. We pass the last hidden state of the BGRU to a linear layer with size equal to the number of target classes (6 for CREMA, 8 for Ravdess, 4 for IEMOCAP) followed by a Softmax layer with a cross entropy loss for emotion classification. This process (self-supervised feature extraction + BGRU training) is followed for all compared methods (see bottom of Fig. \ref{fig:overview}).

For the \textbf{speech recognition} task, we use the SPC dataset to evaluate our method. For the SPC dataset which is a spoken word classification task with 30 different possible labels, we use the exact same protocol as described for emotion recognition (self-supervised feature extraction + BGRU training). We use the same parameters and learning schedule for the BGRU.

For the \textbf{continuous affect recognition} task, we use SEWA and RECOLA as the evaluation datasets. We predict continuous values of valence and arousal for each audio segment. We train separate individual models for valence and arousal prediction. Using the same encoder architecture and self-supervised pretraining followed by downstream evaluation process described in the bottom of Fig. \ref{fig:overview}, we use the output of the BGRU at each timestep as the predicted valence or arousal value. This gives us a predicted sequence $y$. We train the network to maximize the Concordance Correlation Coefficient (CCC) between the predicted sequence $y$ and the ground truth valence or arousal sequence $\hat{y}$. The formula for CCC is:

\begin{equation}
    CCC = \frac{2\sigma_{y\hat{y}}}{\sigma_{y}^{2} + \sigma_{\hat{y}}^{2} + (\mu_{y} - \mu_{\hat{y}})^2}
\end{equation}
where $\mu_{y}$ and $\mu_{\hat{y}}$ are the means of the ground truth and predicted sequences, $\sigma_{y}^{2}$ and $\sigma_{\hat{y}}^{2}$ are the variance of the ground truth and predicted sequences, and $2\sigma_{y\hat{y}}$ is the covariance between the ground truth and predicted sequence. The value of CCC lies between -1 and 1, and is 0 for uncorrelated sequences. The loss function that we minimize to train our network (which maximizes CCC) is 1 - ((CCC+1)/2), which lies between 0 and 1. For SEWA, we report the CCC on the test set as the evaluation metric. For RECOLA, we report the CCC on the validation set to because test set labels are not publicly available. Note that CCC is the standard metric for the SEWA dataset (used in the AVEC 2017 challenge) and the RECOLA dataset (used in the AVEC 2016 challenge).

For all results in Table 2, we validate the models using the validation set of the downstream task and use the model from the epoch with the best validation performance to evaluate on the test set. We report the results in the form of $\mu \pm \sigma$ (mean $\pm$ std) over 10 runs for each experiment.

% For the \textbf{speech recognition} task, we use the GRID and SPC datasets to evaluate our methods. For the SPC dataset which is a spoken word classification task with 30 different possible labels, we use the exact same protocol as described for emotion recognition (self-supervised feature extraction + LSTM training). We use the same parameters and learning schedule for the LSTM.
% However, for the GRID dataset, we have a continuous ASR task instead of classification (i.e. we need to decode the full sentence for every utterance instead of just assigning it a class label). Thus we need to change the evaluation pipeline in order to do WER (word error rate) evaluation instead of classification. For this, we use the extracted features converted to Kaldi format and employ the ESPNet \cite{watanabe2018espnet} toolkit for the end-to-end ASR training. We use a hybrid CTC/attention based ASR model with the default ESPNet parameters with a BLSTM encoder (as used similarly in \cite{petridis2019end}) with 320 units and location aware attention. We train the model for 15 epochs.  For decoding, we use a beam search with a beamsize of 20 and a CTC weight of 0.1.

\subsection{Results with Visual Self-Supervision (L1)}
Our method for visual self-supervision by face reconstruction from audio is based on an L1 reconstruction loss, and is indicated as \textbf{L1} throughout the results in Table \ref{tab:all_results}. We also use a pairwise t-test between observations over 10 runs when discussing the statistical significance of the results throughout Section \ref{sec:results}. For discrete emotion recognition, irrespective of dataset, our method performs better than any audio self-supervised baseline. On CREMA, L1 achieves an F1 score of 0.567, which is significantly better (p$<$0.01) than the best performing baseline PASE, which achieves an F1 score of 0.488. For Ravdess, APC is the best baseline with an F1 score of 0.497, but L1 with an F1 score of 0.560 significantly outperforms this (p$<$0.01). The same trend can be seen for IEMOCAP, with L1 again being the best performing method with an F1 score of 0.618. For speech recognition on SPC, L1 is again the best self-supervised method with an F1 score of 0.935, which outperforms MFCCs (optimised for speech recognition) at 0.912 (significant difference, p$<$0.01). For continuous affect recognition on SEWA, L1 outperforms the other self-supervised methods with a CCC of 0.361 for valence and 0.378 for arousal. The closest baselines to L1 on SEWA are PASE for valence (CCC=0.354, p=0.136) and MFCCs for arousal (CCC=0.359, p=0.051). On RECOLA, L1 again outperforms the other methods with a valence CCC of 0.438 and an arousal CCC of 0.733. The closest competing methods to L1 are again PASE (arousal CCC=0.719, p$<$0.01) and MFCCs (valence CCC=0.413, p$<$0.01).

In summary, the proposed method for visual self-supervision leads to features that significantly outperform those from baseline audio self-supervised methods for both emotion recognition and speech recognition.

\subsection{Results with Audio-only Self-Supervision (Odd)}
Our method for audio-only self-supervision (Odd One Out) is indicated as \textbf{Odd} throughout the discussion of the results.

For discrete emotion recognition, Odd is the best performing audio-only self-supervised method on CREMA (F1 score of 0.542), Ravdess (F1 score of 0.501) and IEMOCAP (F1 score of 0.615). For continuous affect recognition on SEWA, Odd achieves a valence CCC of 0.342, which is outperformed by MFCCs with CCC of 0.349 (p=0.13). For arousal, Odd achieves CCC of 0.339 which is outperformed by both MFCCs and PASE with CCCs of 0.359 (p$<$0.01) and 0.343 (p=0.362) respectively. On RECOLA, Odd is the best method for arousal (CCC=0.720) and valence (CCC=0.436). For speech recognition on SPC, both PASE (F1 score of 0.941) and APC (F1 score of 0.933) significantly (p$<$0.01) outperform Odd (F1 score of 0.919).

We notice that while Odd is the best method seen so far for discrete emotion and offers competitive performance with respect to other methods on the other tasks, there is a performance gap for speech recognition and continuous affect recognition with MFCCs and PASE. We aim to bridge this performance gap and yield better features using a multimodal combination of the proposed methods using mult-task learning.

% For emotion recognition, AoT is the best performing audio-only self-supervised method on CREMA (achieving an emotion recognition accuracy of 48.78\%), while Odd is the best on IEMOCAP (achieving an emotion recognition accuracy of 45.14\%). Both jointly perform the best on Ravdess (AoT: 39.50\%; Odd: 39.49\%). For speech recognition, Odd is the best method both on SPC (89.29\%) and GRID (WER 5.1). PASE is the closest competing self-supervised method for ASR except MFCCs.

% When comparing between the two proposed methods, Odd and AoT seem to be very close in performance on emotion recognition, but Odd seems to slightly outperform AoT on speech recognition (likely due to being a more refined pretext task). Both methods outperform baselines for audio-only self-supervision, however when compared to the L1 method using visual self-supervision, they fall short for all evaluated unimodal experiments. This leads to the observation that the proposed visual self-supervision approach yields better features than all proposed audio-only self-supervised approaches. There is also a performance gap between the proposed unimodal self-supervised methods and MFCCs for speech recognition. We attempt to bridge this gap and yield better features using a multimodal combination of the proposed methods using multi-task learning.

\subsection{Results with Audiovisual Self-Supervision (L1 + Odd)}

% \subsubsection{Optimal multi-task learning weights}
In order to determine the optimal weights for each modality for multi-task learning, we tune the parameter  $ \alpha $ (equation \ref{eq:loss}) on the validation sets of the CREMA, Ravdess and SPC datasets (introduced in section \ref{sec:data}) for a range of values. The results for the tuning are in Table \ref{tab:alphas}. From the table, we observe that the best value of $ \alpha $ is 0.67 for  L1 + Odd. Thus, we use the model trained with this value when evaluating on the test sets in all experiments.

When comparing the results with other results in Table \ref{tab:all_results}, we can see a clear improvement using audiovisual self-supervision. L1 + Odd significantly outperforms all other methods in every experiment. For discrete emotion recognition, L1 + Odd is the best-performing method on CREMA (F1 score = 0.573), IEMOCAP (F1 score = 0.631), and Ravdess (F1 score = 0.565). For speech recognition, L1 + Odd is the best-performing method on SPC (F1 score =  0.944). The notable observation here is that L1 + Odd significantly outperforms either unimodal method in all experiments (p$<$0.01 for all differences from Odd, p$<$0.05 for all differences from L1, except for CREMA with p=0.33 and arousal on SEWA with p=0.37). For speech recognition, this also helps in outperforming PASE (difference with p$<$0.05) and MFCCs (difference with p $<$0.01), which Odd was not able to do on its own. This points to the presence of complementary information being encoded by the two types of supervision from the two modalities which leads to very good generalized audio representations. In summary, the proposed multimodal self-supervision method clearly outperform any proposed unimodal self-supervision method.

\subsection{When does visual self-supervision specifically help?}
We now examine the specific cases and scenarios where visual information is especially helpful to improve the audio representations. For each of the three downstream tasks (discrete emotion, continuous affect, and speech recognition), we compare the class-wise predictions of the model trained using only audio self-supervision (Odd) and that with joint audiovisual self-supervision (L1 + Odd).

\begin{figure}
    \centering
    \includegraphics[width=\columnwidth]{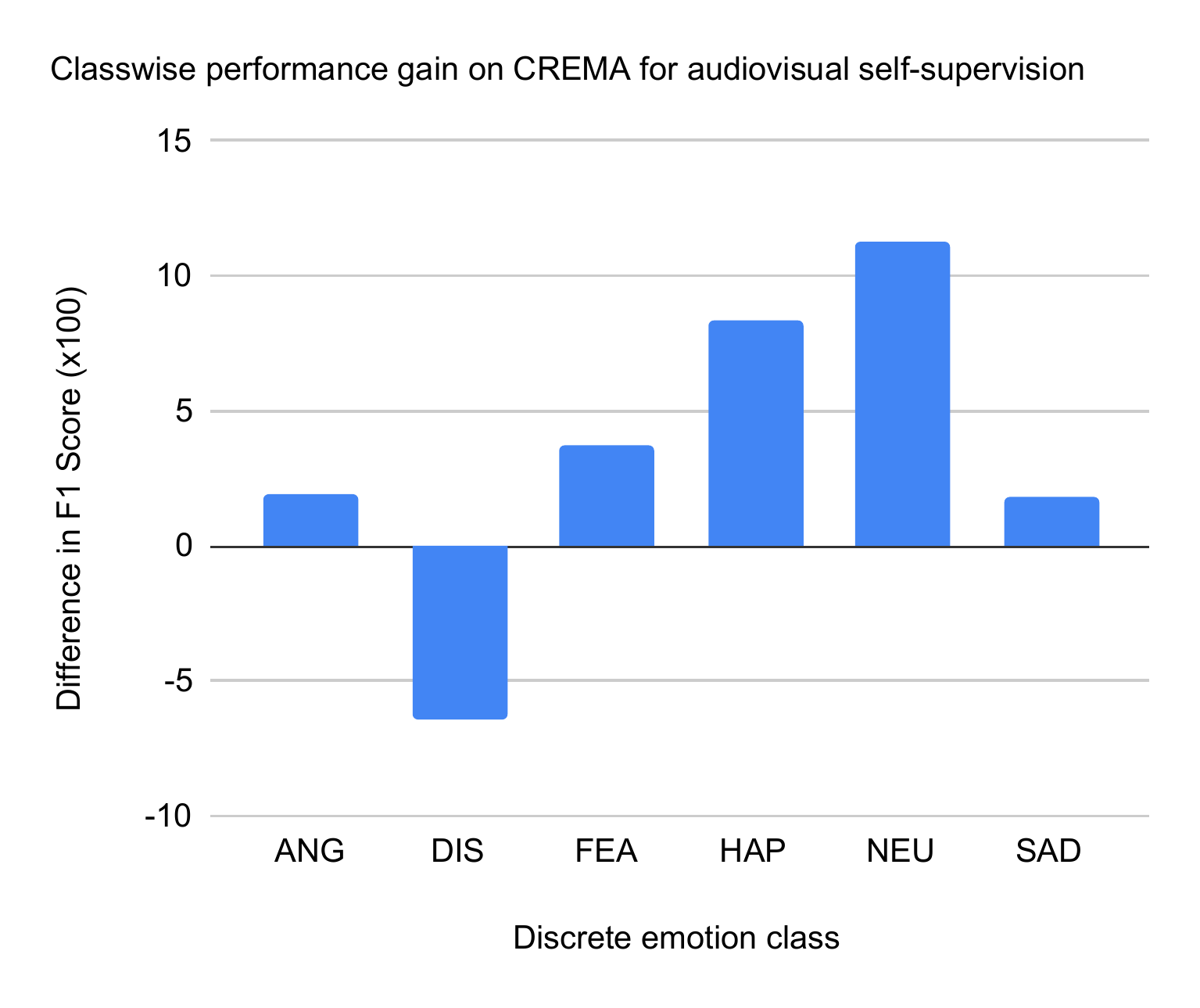}
    \caption{The classwise performance gain of L1 + Odd (audiovisual) compared to Odd (audio-only) on the CREMA dataset. Values are 100x the difference in F1 scores.}
    \label{fig:crema_chart}
\end{figure}

For discrete emotion recognition (see Fig. \ref{fig:crema_chart}), we analyze the difference in performance between the model trained using only audio self-supervision (Odd) and the model trained using joint audiovisual self-supervision (L1 + Odd) for each emotion class in the CREMA dataset. We saw in Table \ref{tab:all_results} that L1 + Odd gets an overall F1 score of 0.573 compared to Odd with F1 score of 0.542. In Fig. \ref{fig:crema_chart}, we see how each class contributes to this performance gain. We find that the gain is the highest for the HAP (happy) and NEU (neutral) classes. There are also moderate gains seen for the ANG (anger), DIS (disgust) and SAD (sadness) classes. However there is a performance drop observed for the DIS (disgust) class. Overall, we conclude that visual information during pretraining helps reduce the confusion between classes such as HAP and NEU (which are more clearly separable in the visual modality than the audio modality). Even though the DIS class is negatively impacted, this results in a significantly better audio representation for discrete emotion.

For continuous affect recognition (see Table \ref{tab:high_low_comparison}), we analyze the difference in performance between Odd (audio self-supervision), L1 (visual self-supervision) and L1 + Odd (audiovisual self-supervision) for different types of audio segments in the RECOLA dataset. We compare the methods for four different types of audio segments (low/high valence and low/high arousal). We use the test CCC attained by Odd as the baseline and compare the performance gain attained with respect to this. We observe that L1 + Odd is the best method for all types of segments. The performance gain from visual information is most pronounced for high valence sequences (+2.15 for L1 and +6.76 for L1 + Odd). We see a very slight gain for low valence sequences (+0.02 for L1 and +0.44 for L1 + Odd). We also see overall gains for both high and low arousal (+1.01/+1.10 for L1 and +2.91/+2.87 for L1 + Odd). Overall, L1 + Odd still offers the best representation for both valence and arousal.

\begin{table}
    \centering
    \renewcommand{\arraystretch}{1.3}
    \begin{tabular}{cccc}
        \toprule
        \textbf{Relative CCC (x100) gain} & \multicolumn{3}{c}{Method}\\
        \cmidrule(lr){1-1} \cmidrule(lr){2-4}
        Type of segment & Audio & Visual & Audiovisual\\
        \cmidrule(lr){1-1} \cmidrule(lr){2-2} \cmidrule(lr){3-3} \cmidrule(lr){4-4}
        Low Valence & 0.00 & +0.02 & +0.44\\
        High Valence & 0.00 & +2.15 & +6.76\\
        % \midrule
        Low Arousal & 0.00 & +1.10 & +2.87\\
        High Arousal & 0.00 & +1.01 & +2.91\\
        \bottomrule
    \end{tabular}
    \caption{Comparison of Odd (Audio), L1 (Visual) and L1 + Odd (Audiovisual) on high/low valence and high/low arousal audio segments from the RECOLA dataset. Performance gains are relative to the CCC attained by Odd on each type of segment. Gain values are CCCx100.}
    \label{tab:high_low_comparison}
\end{table}

\begin{figure}
    \centering
    \includegraphics[width=\columnwidth]{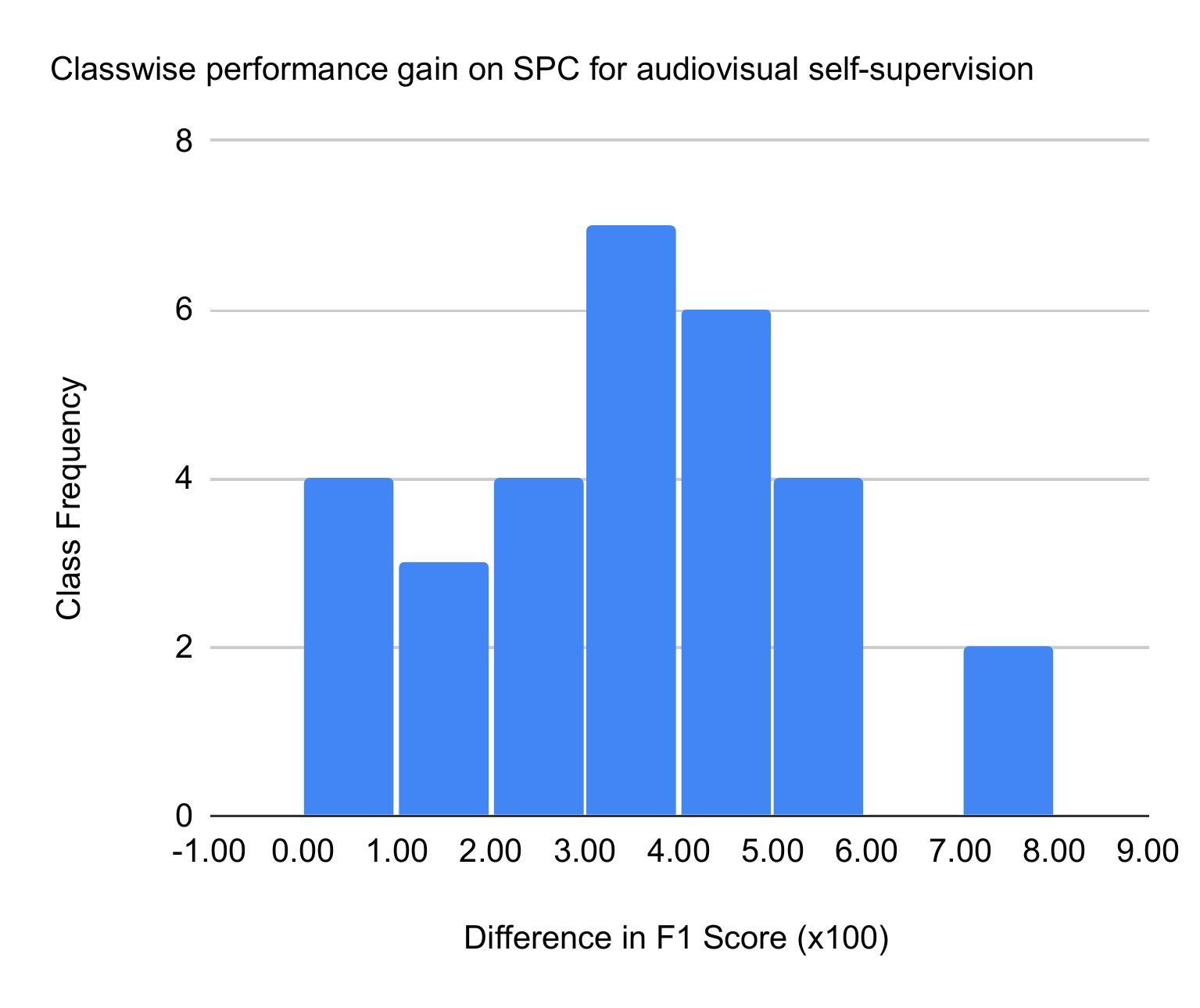}
    \caption{Histogram of the classwise performance gain of L1 + Odd (audiovisual) compared to Odd (audio-only) on the SPC dataset. Values are 100x the difference in F1 scores.}
    \label{fig:spc_chart}
\end{figure}

For speech recognition (see Fig. \ref{fig:spc_chart}), we analyze the class-wise performance difference in performance on the SPC dataset between Odd (audio self-supervision) and L1 + Odd (audiovisual self-supervision). We present a histogram of difference in class-wise performance in Fig. \ref{fig:spc_chart}. We see that most classes are positively influenced with a gain in the range of 2\% to 6\%, with 2 of the 30 classes seeing gains of nearly 8\%. The classes with the highest improvements due to visual information primarily are similar sounding words such as 'no'/'go', 'tree'/'three', and 'bird'/'bed'. All of these classes have similar audio and thus higher intra-class confusion, but have useful distinguishing information in the video modality which our proposed audiovisual self-supervision is able to exploit. While the performance gain is the most significant for these specific classes, we also see an overall performance gain for every class. Thus, L1 + Odd offers the best representation for speech recognition on SPC.

\begin{table*}\centering
\small
\renewcommand{\arraystretch}{1.3}
\begin{tabular}{cccccccc}
\hline
\multicolumn{4}{c}{MTL Weight Tuning for Audio and Visual Tasks} & \multicolumn{2}{c}{\textbf{Emotion Recognition}} & \multicolumn{1}{c}{\textbf{Speech Recognition}}\\ \hline
\multicolumn{4}{c}{Pretraining Dataset} & LRW & LRW & LRW \\
\multicolumn{4}{c}{Evaluation Dataset} & \textbf{CREMA-D} & \textbf{Ravdess} & \textbf{SPC} \\
\multicolumn{4}{c}{Classifier for \textit{(t, dim)} features} & LSTM &  LSTM & LSTM \\
\multicolumn{4}{c}{Labels} & 6 emotions & 8 emotions & 30 words \\\hline
Method & Video weight ($ \alpha $) & Audio weight ($ 1 - \alpha $) & Dim. & Accuracy ($\uparrow$) & Accuracy ($\uparrow$) & Accuracy ($\uparrow$) \\\hline
% L1 + AoT  & 0.17 & 0.83 & 512 & 46.22 & 38.61 & 88.74\\
% L1 + AoT  & 0.33 & 0.67 & 512 & 47.91 & 40.18 & 89.28\\
% L1 + AoT  & 0.50 & 0.50 & 512 & 51.50 & 43.03 & 90.36\\
% L1 + AoT  & \textbf{0.67} & \textbf{0.33} & 512 & \textbf{51.77} & \textbf{44.39} & \textbf{91.94}\\
% L1 + AoT  & 0.83 & 0.17 & 512 & 48.93 & 40.40 & 90.79\\\hline
L1 + Odd  & 0.17 & 0.83 & 512 & 48.91 & 42.11 & 89.97\\
L1 + Odd  & 0.33 & 0.67 & 512 & 47.48 & 39.81 & 88.39\\
L1 + Odd  & 0.50 & 0.50 & 512 & 50.73 & 43.26 & 90.78\\
L1 + Odd  & \textbf{0.67} & \textbf{0.33} & 512 & \textbf{52.81} & \textbf{44.32} & \textbf{92.17}\\
L1 + Odd  & 0.83 & 0.17 & 512 & 51.17 & 42.41 & 91.31\\
\hline
\end{tabular}
\caption{Comparison of different MTL weights. All results are accuracies on the validation sets of the evaluation datasets.}
    \label{tab:alphas}
\end{table*}

\subsection{Performance in various levels of noise}

\begin{figure*}[t]
    \centering
    \subfloat[\scriptsize{Emotion recognition F1 score on CREMA under noise}]{%
       \includegraphics[width=0.5\textwidth]{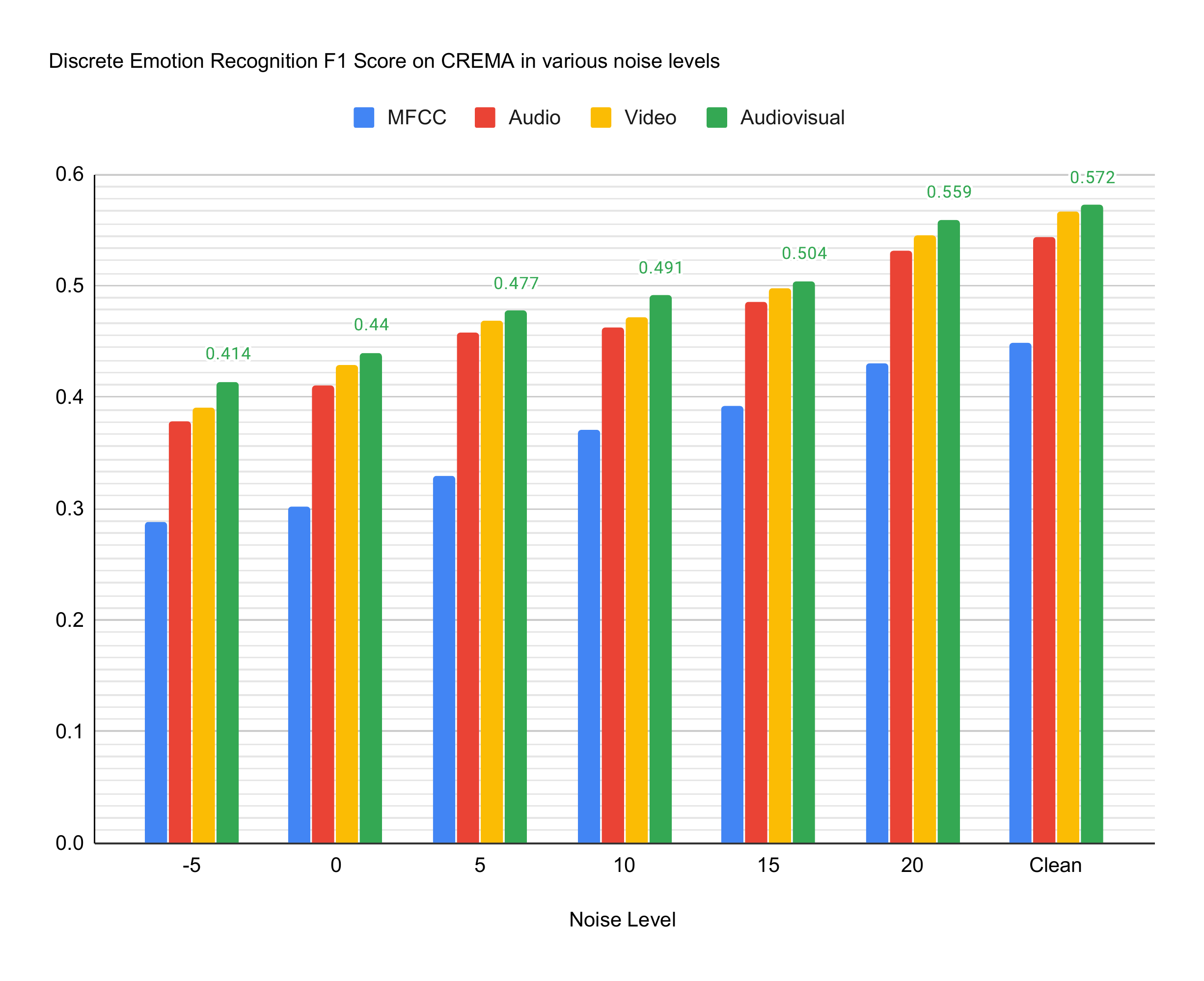}}
   \subfloat[\scriptsize{Speech recognition F1 score on SPC under noise}]{%
       \includegraphics[width=0.5\textwidth]{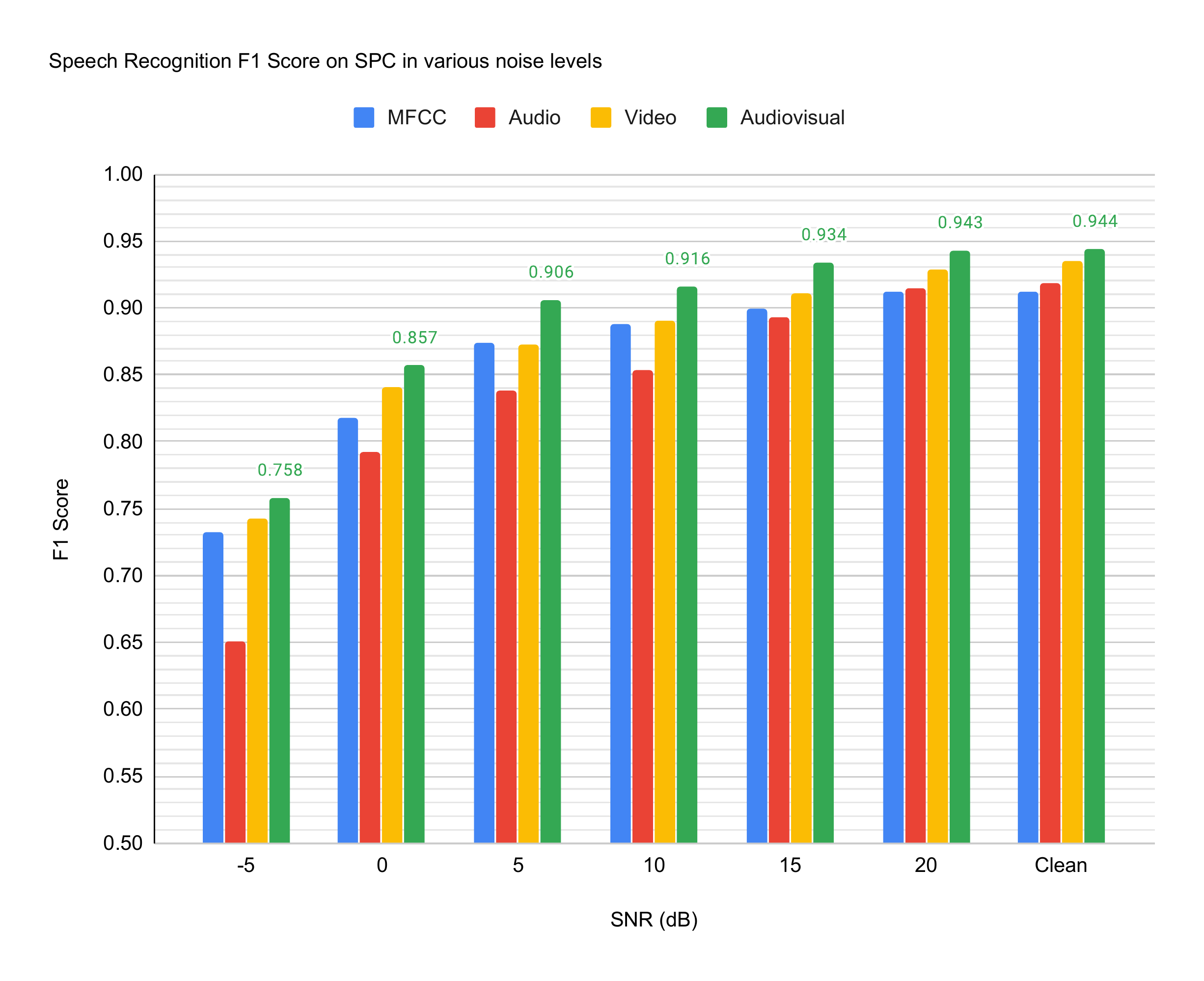}}
    \caption{Comparison of performance of MFCCs and our methods trained using audio-only (Odd), video-only (L1) and audio-visual (L1 + Odd) self-supervision under various levels of artificially introduced babble noise. Best viewed under zoom.}
    \label{fig:noise}
\end{figure*}

% \begin{figure*}[t]
%     \centering
%     \begin{subfigure}{0.5\textwidth}
%     \centering
%     \includesvg[width=\textwidth]{figures/crema-noise.svg}
%     \caption{Emotion recognition accuracy on CREMA under noise}
%     \end{subfigure}%
%     \begin{subfigure}{0.5\textwidth}
%     \centering
%     \includesvg[width=\textwidth]{figures/spc-noise.svg}
%     \caption{Speech recognition accuracy on SPC under noise}
%     \end{subfigure}
%     \caption{Comparison of performance of our methods trained using audio-only (Odd), video-only (L1) and audio-visual (L1 + Odd) self-supervision under various levels of artificially introduced babble noise. Best viewed under zoom.}
%     \label{fig:noise}
% \end{figure*}

In order to further rigorously validate our proposed models for robustness, we investigate the performance under various levels of noise. We create noisy versions of the CREMA and SPC datasets by adding babble noise from the NOISEX database \cite{varga1993assessment}, while varying the SNR from -5 dB to 20 dB in steps of 5 dB. We perform a comparison between the following methods: (i) L1 using visual self-supervision, (ii) Odd using audio-only self-supervision, and (iii) L1 + Odd for bimodal self-supervision, and (iv) MFCCs. We examine how the performance varies for the methods as the level of added noise changes in the evaluation datasets.

\begin{figure*}[t]
    \centering
    \subfloat[\scriptsize{Emotion recognition F1 score on CREMA}]{%
       \includegraphics[width=0.5\textwidth]{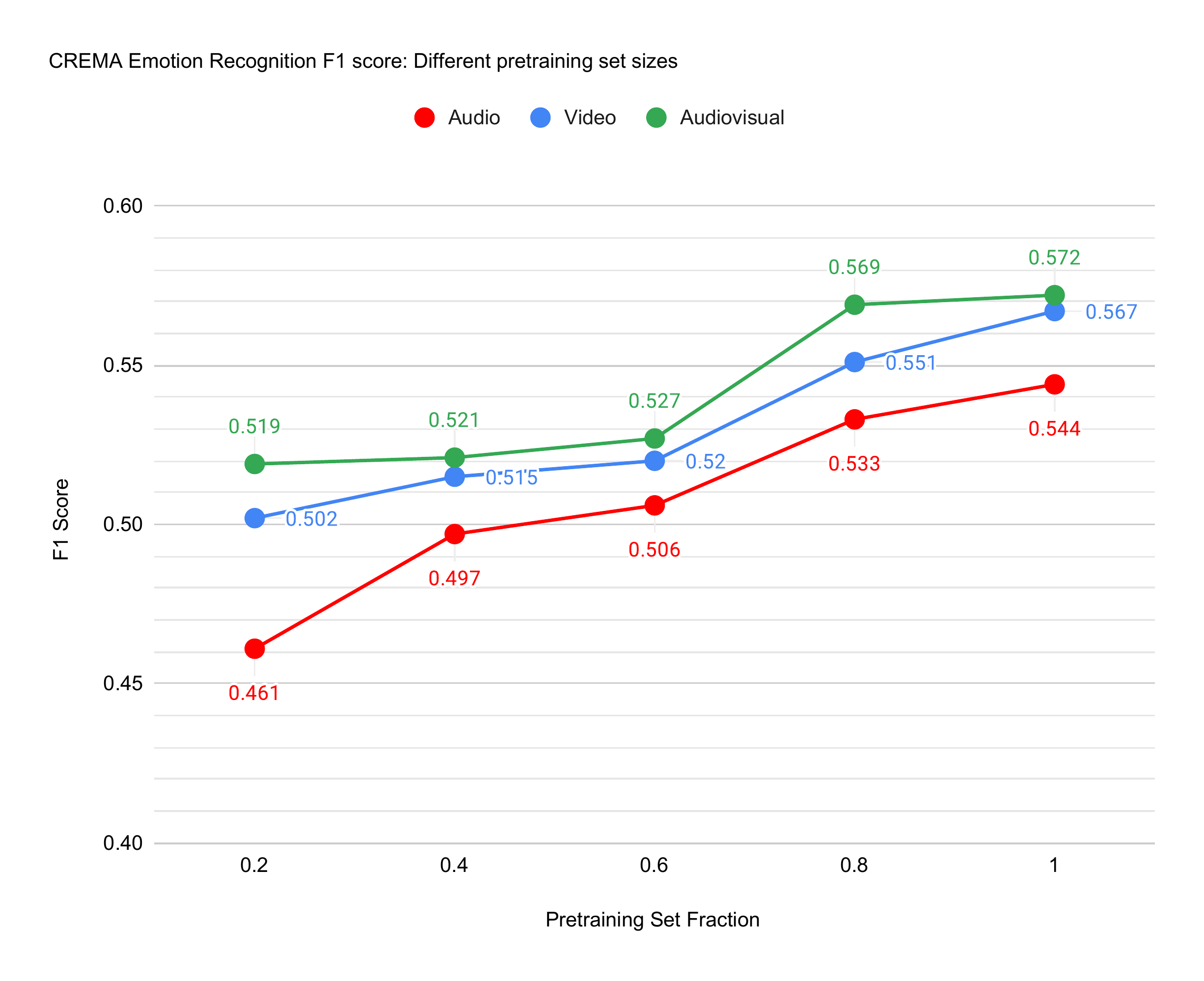}}
   \subfloat[\scriptsize{Speech recognition F1 score on SPC}]{%
       \includegraphics[width=0.5\textwidth]{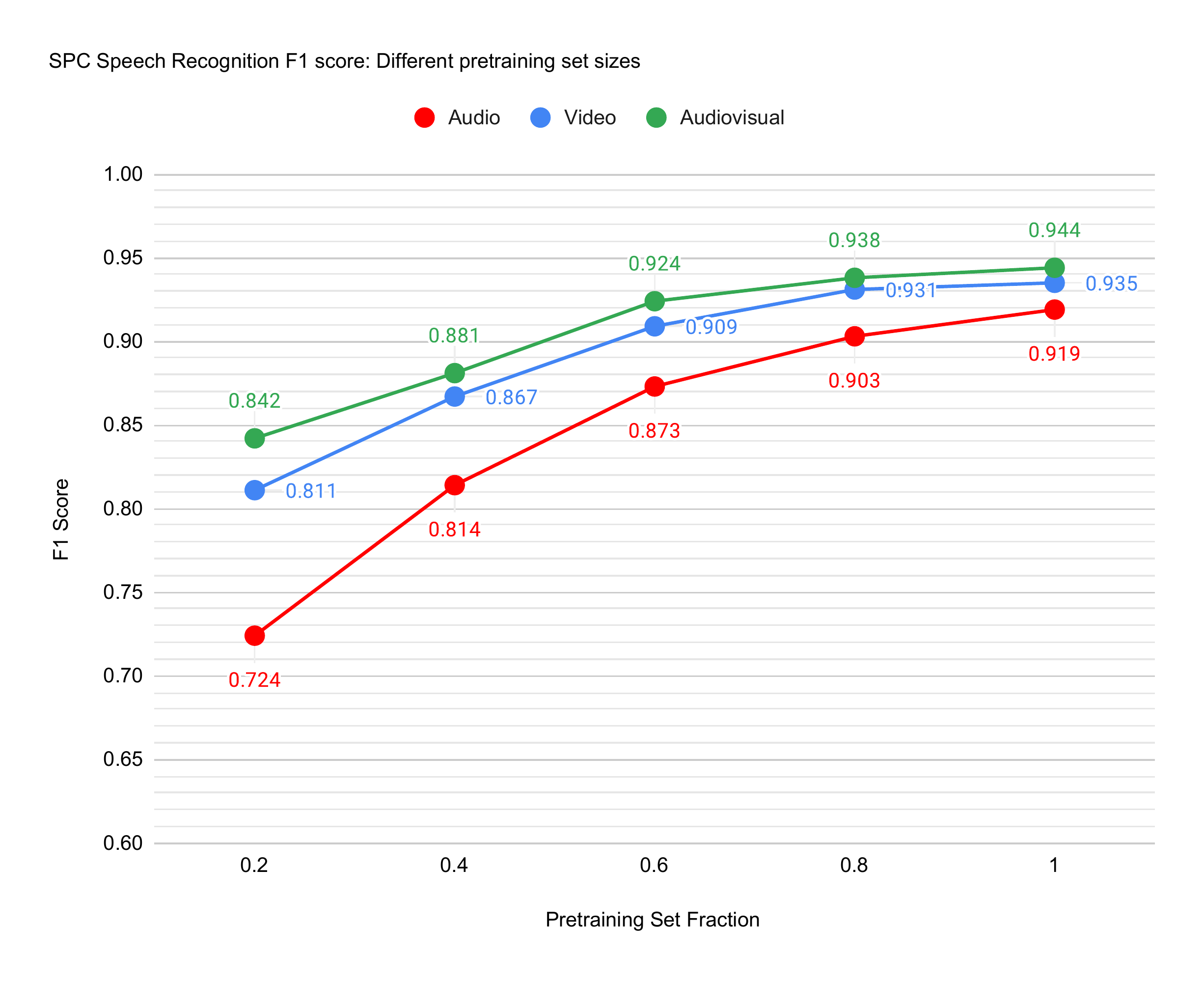}}
    \caption{Comparison of performance of our methods trained using audio-only (Odd), video-only (L1) and audio-visual (L1 + Odd) self-supervision for various sizes of the pretraining set (fraction of total). Total samples in LRW pretraining set = 112658. Best viewed under zoom.}
    \label{fig:pretrainingsize}
\end{figure*}

% \begin{figure*}[t]
%     \centering
%     \begin{subfigure}{0.5\textwidth}
%     \centering
%     \includesvg[width=\textwidth]{figures/crema-pretrainsize.svg}
%     \caption{Emotion recognition accuracy on CREMA}
%     \end{subfigure}%
%     \begin{subfigure}{0.5\textwidth}
%     \centering
%     \includesvg[width=\textwidth]{figures/spc-pretrainsize.svg}
%     \caption{Speech recognition accuracy on SPC}
%     \end{subfigure}
%     \caption{Comparison of performance of our methods trained using audio-only (Odd), video-only (L1) and audio-visual (L1 + Odd) self-supervision for various sizes of the pretraining set (fraction of total). Total samples in LRW pretraining set = 112658. Best viewed under zoom.}
%     \label{fig:pretrainingsize}
% \end{figure*}

The results for the experiments with added noise can be seen in Fig \ref{fig:noise}. For both datasets, we observe that self-supervision using audiovisual combination outperforms unimodal methods. For emotion recognition on CREMA, the audio features from the clean dataset give the best performance, and there is a linear degradation of performance with the increase in noise. Visual self-supervision is more effective in almost all scenarios, which may be expected as visual features are unaffected by auditory noise and can still drive robust learning of audio features. A similar conclusion can be reached for speech recognition as well. Yet audio-visual self-supervision leads to the best audio representations across all noise labels.
% For speech recognition on SPC, the best accuracy is obtained with the SNR of 15 dB (92.46\%). It is slightly surprising that this is better than the clean version of the dataset, however a possible explanation is that a minor level of noise offers some degree of regularization on this particular dataset.
The results for speech recognition are also significantly more robust to noise than those for emotion recognition. We can see this from the decrease in performance with increase in noise, which is not very sharp until extremely high noise levels at -5 dB and 0 dB. We can also notice a sharper degradation for the audio-only results when compared to both the video-only and audiovisual results. This suggests that audio features obtained by visual or audio-visual self-supervision are more robust to noise compared to those obtained by audio-only self-supervision. 

\subsection{Performance with various sizes of the pretraining set and downstream training set}

It is also an interesting experiment to see how model performance varies with the amount of data used for self-supervised pretraining. One of the most important advantages of a self-supervised learning approach is the ability to use an arbitrarily high amount of unlabeled data to learn a good representation. But there is still a tradeoff to be made with training time and model performance. We used a subset of the LRW dataset with a total pretraining size of 112658 samples (36.3 hours of audiovisual speech) to train our full model. For this experiment, we investigate what happens to our model if we only use a fraction of the total available data for pretraining. We use 0.2, 0.4, 0.6 and 0.8 times the dataset size for pretraining. We compare the performance between: (i) L1 for visual self-supervision, (ii) Odd for audio-only self-supervision, and (iii) L1 + Odd for audio-visual self-supervision. As before, we evaluate on the CREMA dataset for emotion recognition and the SPC dataset for speech recognition.

The results for the experiments with different pretraining set sizes can be seen in Fig. \ref{fig:pretrainingsize}. For emotion recognition on CREMA, there is a steady and somewhat linear degradation in performance with the reduction in data in the pretraining. The best performance drops from an F1 score of 0.572 with the full dataset to an F1 score of 0.519 for 20\% of the dataset. For speech recognition on SPC, there is a slower degradation for both visual and audiovisual methods, however there is a sharper degradation for the audio method with lesser training data. With 20\% of the training data, the performance for the audio-only method drops to an F1 score of 0.724, which is a massive gap from the F1 score of 0.842 for the audiovisual method in the same setting. This goes on to show that the method with combined audio and visual self-supervision offers more robustness with varied amounts of pretraining data. Another observation is that a larger amount of pretraining data helps learn better features: the gain is significant initially however starts to plateau after a point (80\% of the data).
Fig. \ref{fig:pretrainingsize} also shows that with just 20\% / 80\% of the LRW dataset for pretraining, the audio-visual self-supervised methods achieve similar or better performance ($\pm$0.5 for emotion, $\pm$0.94 for ASR) as PASE, CPC and APC which use the full pretraining set (see Table \ref{tab:all_results}). This is a very interesting result because it shows that our proposed self-supervised methods require lesser pretraining data than other tested self-supervised methods to achieve competitive performance.

We also perform an experiment on SPC in which we simulate a limited labeled data setting by using only 10\% of the training set labels for the downstream tasks. We present the results for this experiment in Table \ref{tab:spc_10}. We can see that L1 + Odd is the best performing method in the limited data setting with an F1 score of 0.897. We also notice a much sharper degradation in the performance of other baseline methods as compared to our visual and audiovisual self-supervision based methods. Our method being the best performing method in the limited data setting is also very relevant in the context of self-supervision, where the goal is to use a large unlabeled dataset to learn useful representations which are then evaluated on a smaller labeled dataset.

\begin{table}
    \centering
    \begin{tabular}{ccc}
    \toprule
       \textbf{SPC: Limited label setting}  &  \multicolumn{2}{c}{\% of labels used}\\
    \cmidrule(lr){1-1} \cmidrule(lr){2-3}
    Method & 100\% & 10\%\\
    \cmidrule(lr){1-1} \cmidrule(lr){2-2} \cmidrule(lr){3-3}
        MFCC & 0.912 $\pm$ 0.002 &  0.864 $\pm$ 0.013\\
        PASE & 0.941 $\pm$ 0.002 & 0.811 $\pm$ 0.025 \\
        APC & 0.933 $\pm$ 0.003 & 0.874 $\pm$ 0.011 \\
        Odd & 0.919 $\pm$ 0.002 & 0.839 $\pm$ 0.019 \\
        L1 & 0.935 $\pm$ 0.003 & 0.877 $\pm$ 0.012 \\
        L1 + Odd & \textbf{0.944 $\pm$ 0.002} & \textbf{0.897 $\pm$ 0.008} \\
    \bottomrule
    \end{tabular}
    \caption{Comparison of speech recognition performance on SPC with full training set used and with 10\% of the training set used to simulate a limited labeled data setting. Results are F1 scores.}
    \label{tab:spc_10}
\end{table}

% It would be an interesting experiment to pretrain with even larger datasets to see just how much model performance can be improved before the amount of data does not matter anymore.

\subsection{Comparison with finetuned and supervised versions of encoders}
All the results discussed thus far were with the frozen versions of the encoders, i.e. the encoders with their fixed weights were used as feature extractors on the evaluation datasets before training a classifier. However, these encoders can also be fine-tuned to the target dataset. We also present results for finetuning in the bottom of Table \ref{tab:all_results}. We use the weights from the L1 + Odd model as the initialization for the encoder before performing training on the target datasets in an end-to-end manner. We observe that finetuning is able to give us better results depending on the problem setting. The best result that we get on CREMA is an F1 score of 0.592, which is significantly better (p$<$0.05) than the F1 score of 0.573 that is attained using just the frozen encoder. For SPC, we get an F1 score of 0.953 which is also the best result seen so far in all experiments. The largest observed gain comes for the Ravdess dataset, on which we get an F1 score of 0.645, compared to the F1 score of 0.565 with the frozen version. We also see gains using when finetuning on SEWA and RECOLA for continuous affect recognition. For SEWA, the valence CCC improves from 0.373 to 0.380 and the arousal CCC improves from 0.371 to 0.383. For RECOLA, the valence CCC improves from 0.448 to 0.452 and the arousal CCC improves from 0.745 to 0.764. All improvements on SEWA and RECOLA are significant with p$<$0.05.

% \begin{figure*}[t]
%     \centering
%     \subfloat[\scriptsize{Training loss per epoch}]{%
%       \includegraphics[width=0.5\textwidth]{figures/trainloss.png}}
%   \subfloat[\scriptsize{Validation accuracy per epoch}]{%
%       \includegraphics[width=0.5\textwidth]{figures/valcurve.png}}
%     \caption{Learning curves for the finetuned and supervised versions of the models when evaluating on the Ravdess dataset for emotion recognition. It is clear that the finetuned versions of the model start from a superior weight initialization and offer faster convergence. Finetuning also provides better overall performance with half the epochs.}
%     \label{fig:curves}
% \end{figure*}

% \begin{figure*}[t]
%     \centering
%     \begin{subfigure}{0.5\textwidth}
%     \centering
%     \includegraphics[width=\textwidth]{figures/trainloss.png}
%     \caption{Training loss per epoch}
%     \end{subfigure}%
%     \begin{subfigure}{0.5\textwidth}
%     \centering
%     \includegraphics[width=\textwidth]{figures/valcurve.png}
%     \caption{Validation accuracy per epoch}
%     \end{subfigure}
%     \caption{Learning curves for the finetuned and supervised versions of the models when evaluating on the Ravdess dataset for emotion recognition. It is clear that the finetuned versions of the model start from a superior weight initialization and offer faster convergence. Finetuning also provides better overall performance with half the epochs.}
%     \label{fig:curves}
% \end{figure*}

To motivate the utility of adopting self-supervised pretraining, it is also important to compare our method with a supervised version of an encoder with the same architecture trained from scratch directly on the target dataset (see bottom of Table \ref{tab:all_results}). The only difference is the weight initialization, for which we use random initialization for each layer. We observe that our proposed method (L1 + Odd) significantly outperforms (p$<$0.01) the fully supervised method on every task. This is an important result, highlighting that a large amount of unlabeled pretraining data can be used to improve the performance on small labeled downstream datasets.

% It is also interesting to compare our methods with a supervised version of an encoder with the same architecture trained from scratch directly on the target dataset (see bottom of Table \ref{tab:all_results}). The only difference is the weight initialization, for which we use random initialization for each layer. We compare the supervised version to the finetuned version for the exact same hyperparameter sets (Enc LR and Cls LR). We find that for the first two parameter sets, the supervised model is not able to learn any useful features at all and attains performance close to chance. For the other parameter sets with Enc LR = 10e-4, the supervised model does converge and offers good performance. However this is still significantly worse than that obtained by the finetuned version (see \ref{tab:finetuned}). This is clear evidence to support that our self-supervised pretraining yields a much better weight initialization that is likely to converge for a wider variety of hyperparameters while training on downstream tasks that have smaller datasets.

% We observe from Fig. \ref{fig:curves} that training a fully supervised model from scratch without self-supervised pretraining is more susceptible to overfitting and non-convergence for certain hyperparameters. It also results in significantly slower training (as can be seen from Fig. \ref{fig:curves}). We are able to achieve much better performance by finetuning our model for every tested parameter set, despite training for only half the number of epochs.

\section{Discussion\label{sec:discussion}}
There are a number of interesting observations from the experiments. 

The audio-only self-supervised method offers competitive performance but does not always outperform the existing audio self-supervised baselines. There is also a clear observation that the proposed visual self-supervision is superior when compared to the proposed audio-only self-supervision. This is largely due to the fact that the audio features obtained by visual self-supervision are closely related to the useful information present in lip movements and facial expressions (because they must encode this information for accurate facial reconstruction during pretraining). This property is also especially useful for emotion recognition due to the correlation between emotion and facial expression information, and for speech recognition due to the information from lip movements. The results on both discrete and continuous affect recognition offer evidence that the learned representation is good for emotion.

It is also clear that the models trained using a combination of audio and visual self-supervision are able to encode complementary information from each modality to yield the best possible representations among all tested methods in this work. These representations are also the most robust in the presence of various levels of noise in the data, and offer the best performance independently of the size of the pretraining set. This is perhaps the most useful finding of this work, with the implication being that problems using audio-only speech data can benefit from using visual supervision from available audiovisual speech datasets to enhance the target representations (i.e. multimodal pretraining can be used effectively for unimodal downstream evaluation).

Another very useful finding of the work is that fine-tuning of the pretrained audiovisual self-supervised models offers significantly better performance than fully supervised training from scratch. This highlights the applicability of self-supervised pretraining in the area of affective computing. The possibility of learning more generalizable features from large unlabeled multimodal datasets before evaluating on small labeled datasets is particularly promising.
% but faster training and convergence for a variety of hyperparameters when compared to training a fully supervised model from scratch. This could be useful in setting a strong baseline for other speech related problems and cutting down training time on downstream tasks on small datasets, which is a typical problem setting in various domains.
We will make our pretrained models publicly available as to enable the community to commence further research in the area.
% which we hope will be of use to the community.

\subsection{Limitations}
A current limitation to our work is the fact that we use a nearly-frontal subset of the LRW dataset (with yaw, pitch and roll restricted to a maximum of 10 degrees each) for pretraining. This leaves out a large portion of the audiovisual dataset with profile faces which could also contain useful visual supervisory signals. There are also other larger datasets like AVSpeech \cite{ephrat2018looking} which could potentially yield better pretrained models. Another limitation is that we have used a very simple 2 layer BGRU with 256 hidden units as the classifier of choice for our audio classification tasks. This might not be the most optimal method or configuration. However this was chosen for simplicity. Other models such as LSTMs, Transformers, LiGRUs \cite{ravanelli2018light} or temporal convolutional networks (TCNs) in different configurations may yield even better results. Another possible limitation is the fact that the models in our work start from log mel spectrograms instead of raw audio (as input to the audio encoder). There is a static frequency domain transformation applied to raw audio to yield the spectrogram representation, however a more refined approach might be to use a set of trainable filters (e.g. as used in SincNet \cite{ravanelli2018interpretable}) instead of static Mel filters. Another specific limitation in our experiments is that we do not account for annotation delay or do any post-processing for continuous affect recognition, both of which can potentially improve the recognition performance. In summary, the methods presented in this paper show the principle that visual and bimodal self-supervision lead to much better performances than full supervision from scratch. However, more refined approaches (that are tailored to the specific problems and datasets they are developed for) may result in even better performances than those presented here.

\subsection{Future work}
In this work, we have considered the interaction between the audio and visual modalities and how visual self-supervision can benefit learning of audio features. There are also other modalities that could be considered, especially the text modality. Multimodal human language is comprised of text, audio and video combined, and developing a self-supervised model that can capture the interactions between the three can be very useful.
The visual pretext task that we focused on was facial reconstruction optimized by the L1 loss. This process leads to a very realistic facial animation, however this might not be the most desirable thing in order to learn the best features. Realistic reconstruction will need to capture a lot of additional information related to fine grained visual characteristics. A lot of this information might not be useful if our end goal is simply to learn useful audio representations. Although reconstruction does give us really good performance, the question remains open to what a good alternative or additional visual pretext task might be.
This work has also focused on audio features alone. It is also interesting to see how we could use audio self-supervision to guide the learning of visual speech features in an analogous way (by predicting the audio waveform from only the visual modality, like in \cite{vougioukas2019video}). These visual features could then be used for problems like facial affect recognition and lipreading, or even combined with our proposed audio features. There is also a promising and interesting line of work developing in audiovisual contrastive self-supervision (such as in \cite{patrick2020multi, ma2020learning}), which has been demonstrated on downstream tasks like video action recognition and acoustic event detection. It would be interesting to see how such approaches perform when pretrained on audiovisual speech data and applied on emotion recognition.

\ifCLASSOPTIONcompsoc
  \section*{Acknowledgments}
\else
  \section*{Acknowledgment}
\fi
All datasets used in the experiments and all training, testing and ablation studies have been conducted at Imperial College London. Abhinav Shukla's work was supported by a PhD scholarship from Samsung Electronics UK.

\ifCLASSOPTIONcaptionsoff
  \newpage
\fi



%\bibliographystyle{IEEEtran}
\bibliographystyle{IEEEtran}
% \footnotesize{
\bibliography{refs}

\begin{thebibliography}{10}
\providecommand{\url}[1]{#1}
\csname url@samestyle\endcsname
\providecommand{\newblock}{\relax}
\providecommand{\bibinfo}[2]{#2}
\providecommand{\BIBentrySTDinterwordspacing}{\spaceskip=0pt\relax}
\providecommand{\BIBentryALTinterwordstretchfactor}{4}
\providecommand{\BIBentryALTinterwordspacing}{\spaceskip=\fontdimen2\font plus
\BIBentryALTinterwordstretchfactor\fontdimen3\font minus
  \fontdimen4\font\relax}
\providecommand{\BIBforeignlanguage}[2]{{%
\expandafter\ifx\csname l@#1\endcsname\relax
\typeout{** WARNING: IEEEtran.bst: No hyphenation pattern has been}%
\typeout{** loaded for the language `#1'. Using the pattern for}%
\typeout{** the default language instead.}%
\else
\language=\csname l@#1\endcsname
\fi
#2}}
\providecommand{\BIBdecl}{\relax}
\BIBdecl

\bibitem{alwassel2019self}
H.~Alwassel, D.~Mahajan, L.~Torresani, B.~Ghanem, and D.~Tran,
  ``Self-supervised learning by cross-modal audio-video clustering,''
  \emph{arXiv preprint arXiv:1911.12667}, 2019.

\bibitem{peters2018deep}
M.~Peters, M.~Neumann, M.~Iyyer, M.~Gardner, C.~Clark, K.~Lee, and
  L.~Zettlemoyer, ``Deep contextualized word representations,'' in
  \emph{NAACL-HLT}, 2018, pp. 2227--2237.

\bibitem{devlin2018bert}
J.~Devlin, M.-W. Chang, K.~Lee, and K.~Toutanova, ``Bert: Pre-training of deep
  bidirectional transformers for language understanding,'' in \emph{NAACL-HLT},
  2019, pp. 4171--4186.

\bibitem{gidaris2018unsupervised}
S.~Gidaris, P.~Singh, and N.~Komodakis, ``Unsupervised representation learning
  by predicting image rotations,'' in \emph{ICLR}, 2018.

\bibitem{doersch2015unsupervised}
C.~Doersch, A.~Gupta, and A.~Efros, ``Unsupervised visual representation
  learning by context prediction,'' in \emph{ICCV}, 2015, pp. 1422--1430.

\bibitem{fernando2017self}
B.~Fernando, H.~Bilen, E.~Gavves, and S.~Gould, ``Self-supervised video
  representation learning with odd-one-out networks,'' in \emph{CVPR}, 2017,
  pp. 3636--3645.

\bibitem{caron2020unsupervised}
M.~Caron, I.~Misra, J.~Mairal, P.~Goyal, P.~Bojanowski, and A.~Joulin,
  ``Unsupervised learning of visual features by contrasting cluster
  assignments,'' \emph{arXiv preprint arXiv:2006.09882}, 2020.

\bibitem{he2019momentum}
K.~He, H.~Fan, Y.~Wu, S.~Xie, and R.~Girshick, ``Momentum contrast for
  unsupervised visual representation learning,'' in \emph{CVPR}, 2020, pp.
  9729--9738.

\bibitem{misra2019self}
I.~Misra and L.~v.~d. Maaten, ``Self-supervised learning of pretext-invariant
  representations,'' in \emph{CVPR}, 2020, pp. 6707--6717.

\bibitem{donahue2019large}
J.~Donahue and K.~Simonyan, ``Large scale adversarial representation
  learning,'' in \emph{NeurIPS}, 2019, pp. 10\,541--10\,551.

\bibitem{wang2019learning}
X.~Wang, A.~Jabri, and A.~A. Efros, ``Learning correspondence from the
  cycle-consistency of time,'' in \emph{CVPR}, 2019, pp. 2566--2576.

\bibitem{caron2018deep}
M.~Caron, P.~Bojanowski, A.~Joulin, and M.~Douze, ``Deep clustering for
  unsupervised learning of visual features,'' in \emph{ECCV}, 2018, pp.
  132--149.

\bibitem{xie2019self}
Q.~Xie, M.-T. Luong, E.~Hovy, and Q.~V. Le, ``Self-training with noisy student
  improves imagenet classification,'' in \emph{CVPR}, 2020, pp.
  10\,687--10\,698.

\bibitem{zhai2019s4l}
X.~Zhai, A.~Oliver, A.~Kolesnikov, and L.~Beyer, ``S4l: Self-supervised
  semi-supervised learning,'' in \emph{ICCV}, 2019, pp. 1476--1485.

\bibitem{kolesnikov2019revisiting}
A.~Kolesnikov, X.~Zhai, and L.~Beyer, ``Revisiting self-supervised visual
  representation learning,'' in \emph{CVPR}, 2019, pp. 1920--1929.

\bibitem{oord2018representation}
A.~Oord, Y.~Li, and O.~Vinyals, ``Representation learning with contrastive
  predictive coding,'' \emph{arXiv preprint arXiv:1807.03748}, 2018.

\bibitem{chung2019unsupervised}
Y.~Chung, W.~Hsu, H.~Tang, and J.~Glass, ``An unsupervised autoregressive model
  for speech representation learning,'' \emph{Interspeech}, 2019.

\bibitem{ravanelli2018learning}
M.~Ravanelli and Y.~Bengio, ``Learning speaker representations with mutual
  information,'' \emph{Interspeech}, 2019.

\bibitem{schneider2019wav2vec}
S.~Schneider, A.~Baevski, R.~Collobert, and M.~Auli, ``wav2vec: Unsupervised
  pre-training for speech recognition,'' \emph{Interspeech}, 2019.

\bibitem{tagliasacchi2019self}
M.~Tagliasacchi, B.~Gfeller, F.~Quitry, and D.~Roblek, ``Self-supervised audio
  representation learning for mobile devices,'' \emph{arXiv:1905.11796}, 2019.

\bibitem{pascual2019learning}
S.~Pascual, M.~Ravanelli, J.~Serr{\`a}, A.~Bonafonte, and Y.~Bengio, ``Learning
  problem-agnostic speech representations from multiple self-supervised
  tasks,'' \emph{Interspeech}, 2019.

\bibitem{kumar2019secost}
A.~Kumar and V.~K. Ithapu, ``Secost:: Sequential co-supervision for large scale
  weakly labeled audio event detection,'' in \emph{ICASSP 2020}.\hskip 1em plus
  0.5em minus 0.4em\relax IEEE, 2020, pp. 666--670.

\bibitem{quitry2019learning}
F.~d.~C. Quitry, M.~Tagliasacchi, and D.~Roblek, ``Learning audio
  representations via phase prediction,'' \emph{arXiv preprint
  arXiv:1910.11910}, 2019.

\bibitem{chorowski2019unsupervised}
J.~Chorowski, R.~J. Weiss, S.~Bengio, and A.~van~den Oord, ``Unsupervised
  speech representation learning using wavenet autoencoders,'' \emph{IEEE/ACM
  transactions on audio, speech, and language processing}, vol.~27, no.~12, pp.
  2041--2053, 2019.

\bibitem{riviere2020unsupervised}
M.~Rivi{\`e}re, A.~Joulin, P.-E. Mazar{\'e}, and E.~Dupoux, ``Unsupervised
  pretraining transfers well across languages,'' \emph{ICASSP}, 2020.

\bibitem{chung2016lip}
J.~Chung and A.~Zisserman, ``Lip reading in the wild,'' in \emph{ACCV}, 2016.

\bibitem{multisensory2018}
A.~Owens and A.~Efros, ``Audio-visual scene analysis with self-supervised
  multisensory features,'' \emph{ECCV}, 2018.

\bibitem{pham2019found}
H.~Pham, P.~Liang, T.~Manzini, L.~Morency, and B.~P{\'o}czos, ``Found in
  translation: Learning robust joint representations by cyclic translations
  between modalities,'' in \emph{AAAI}, vol.~33, 2019, pp. 6892--6899.

\bibitem{owens2018learning}
A.~Owens, J.~Wu, J.~McDermott, W.~Freeman, and A.~Torralba, ``Learning sight
  from sound: Ambient sound provides supervision for visual learning,''
  \emph{IJCV}, vol. 126, no.~10, pp. 1120--1137, 2018.

\bibitem{petridis2015prediction}
S.~Petridis and M.~Pantic, ``Prediction-based audiovisual fusion for
  classification of non-linguistic vocalisations,'' \emph{IEEE Transactions on
  Affective Computing}, vol.~7, no.~1, pp. 45--58, 2015.

\bibitem{piergiovanni2020evolving}
A.~Piergiovanni, A.~Angelova, and M.~S. Ryoo, ``Evolving losses for
  unsupervised video representation learning,'' \emph{CVPR}, 2020.

\bibitem{tian2019contrastive}
Y.~Tian, D.~Krishnan, and P.~Isola, ``Contrastive multiview coding,''
  \emph{arXiv preprint arXiv:1906.05849}, 2019.

\bibitem{patrick2020multi}
M.~Patrick, Y.~M. Asano, R.~Fong, J.~F. Henriques, G.~Zweig, and A.~Vedaldi,
  ``Multi-modal self-supervision from generalized data transformations,''
  \emph{arXiv preprint arXiv:2003.04298}, 2020.

\bibitem{morgado2020audio}
P.~Morgado, N.~Vasconcelos, and I.~Misra, ``Audio-visual instance
  discrimination with cross-modal agreement,'' \emph{arXiv preprint
  arXiv:2004.12943}, 2020.

\bibitem{zhu2020deep}
H.~Zhu, M.~Luo, R.~Wang, A.~Zheng, and R.~He, ``Deep audio-visual learning: A
  survey,'' \emph{arXiv preprint arXiv:2001.04758}, 2020.

\bibitem{bregler1997video}
C.~Bregler, M.~Covell, and M.~Slaney, ``Video rewrite: Driving visual speech
  with audio,'' in \emph{Proceedings of the 24th annual conference on Computer
  graphics and interactive techniques}, 1997, pp. 353--360.

\bibitem{ezzat2002trainable}
T.~Ezzat, G.~Geiger, and T.~Poggio, ``Trainable videorealistic speech
  animation,'' \emph{ACM Transactions on Graphics (TOG)}, vol.~21, no.~3, pp.
  388--398, 2002.

\bibitem{suwajanakorn2017synthesizing}
S.~Suwajanakorn, S.~M. Seitz, and I.~Kemelmacher-Shlizerman, ``Synthesizing
  obama: learning lip sync from audio,'' \emph{ACM Transactions on Graphics
  (TOG)}, vol.~36, no.~4, pp. 1--13, 2017.

\bibitem{mcgurk1976hearing}
H.~McGurk and J.~MacDonald, ``Hearing lips and seeing voices,'' \emph{Nature},
  vol. 264, no. 5588, p. 746, 1976.

\bibitem{petridis2018end}
S.~Petridis, T.~Stafylakis, P.~Ma, F.~Cai, G.~Tzimiropoulos, and M.~Pantic,
  ``End-to-end audiovisual speech recognition,'' in \emph{2018 IEEE
  International Conference on Acoustics, Speech and Signal Processing
  (ICASSP)}.\hskip 1em plus 0.5em minus 0.4em\relax IEEE, 2018, pp. 6548--6552.

\bibitem{cohn2009detecting}
J.~F. Cohn, T.~S. Kruez, I.~Matthews, Y.~Yang, M.~H. Nguyen, M.~T. Padilla,
  F.~Zhou, and F.~De~la Torre, ``Detecting depression from facial actions and
  vocal prosody,'' in \emph{2009 3rd International Conference on Affective
  Computing and Intelligent Interaction and Workshops}.\hskip 1em plus 0.5em
  minus 0.4em\relax IEEE, 2009, pp. 1--7.

\bibitem{shukla2017evaluating}
A.~Shukla, S.~S. Gullapuram, H.~Katti, K.~Yadati, M.~Kankanhalli, and
  R.~Subramanian, ``Evaluating content-centric vs. user-centric ad affect
  recognition,'' in \emph{Proceedings of the 19th ACM International Conference
  on Multimodal Interaction}, 2017, pp. 402--410.

\bibitem{shukla2018looking}
A.~Shukla, H.~Katti, M.~Kankanhalli, and R.~Subramanian, ``Looking beyond a
  clever narrative: Visual context and attention are primary drivers of affect
  in video advertisements,'' in \emph{Proceedings of the 20th ACM International
  Conference on Multimodal Interaction}, 2018, pp. 210--219.

\bibitem{shukla2020recognition}
A.~Shukla, S.~S. Gullapuram, H.~Katti, M.~Kankanhalli, S.~Winkler, and
  R.~Subramanian, ``Recognition of advertisement emotions with application to
  computational advertising,'' \emph{IEEE Transactions on Affective Computing},
  2020.

\bibitem{kollias2019deep}
D.~Kollias, P.~Tzirakis, M.~A. Nicolaou, A.~Papaioannou, G.~Zhao, B.~Schuller,
  I.~Kotsia, and S.~Zafeiriou, ``Deep affect prediction in-the-wild: Aff-wild
  database and challenge, deep architectures, and beyond,'' \emph{International
  Journal of Computer Vision}, pp. 1--23, 2019.

\bibitem{zadeh2018multimodal}
A.~B. Zadeh, P.~P. Liang, S.~Poria, E.~Cambria, and L.-P. Morency, ``Multimodal
  language analysis in the wild: Cmu-mosei dataset and interpretable dynamic
  fusion graph,'' in \emph{Proceedings of the 56th Annual Meeting of the
  Association for Computational Linguistics (Volume 1: Long Papers)}, 2018, pp.
  2236--2246.

\bibitem{shukla2017affect}
A.~Shukla, S.~S. Gullapuram, H.~Katti, K.~Yadati, M.~Kankanhalli, and
  R.~Subramanian, ``Affect recognition in ads with application to computational
  advertising,'' in \emph{Proceedings of the 25th ACM international conference
  on Multimedia}, 2017, pp. 1148--1156.

\bibitem{vougioukas2018end}
K.~Vougioukas, S.~Petridis, and M.~Pantic, ``End-to-end speech-driven facial
  animation with temporal gans,'' \emph{Proceedings of the British Conference
  on Machine Vision (BMVC)}, 2018.

\bibitem{shukla2020visually}
A.~Shukla, K.~Vougioukas, P.~Ma, S.~Petridis, and M.~Pantic, ``Visually guided
  self supervised learning of speech representations,'' \emph{Proceedings of
  the International Conference on Acoustics Speech and Signal Processing
  (ICASSP)}, 2020.

\bibitem{shukla2020cvprw}
A.~Shukla, S.~Petridis, and M.~Pantic, ``Visual self-supervision by facial
  reconstruction for speech representation learning,'' \emph{Sight and Sound
  Workshop, CVPR}, 2020.

\bibitem{shukla2020icmlw}
A.~Shukla, S.~Petridis, and M.~Pantic, ``Learning speech representations from
  raw audio by joint audiovisual self-supervision,'' \emph{Workshop on
  Self-supervision in Audio and Speech, ICML}, 2020.

\bibitem{ronneberger2015u}
O.~Ronneberger, P.~Fischer, and T.~Brox, ``U-net: Convolutional networks for
  biomedical image segmentation,'' in \emph{MICCAI}, 2015, pp. 234--241.

\bibitem{vougioukas2019realistic}
K.~Vougioukas, S.~Petridis, and M.~Pantic, ``Realistic speech-driven facial
  animation with gans,'' \emph{International Journal of Computer Vision}, pp.
  1--16, 2019.

\bibitem{pytorch}
A.~Paszke, S.~Gross, F.~Massa, A.~Lerer, J.~Bradbury, G.~Chanan, T.~Killeen,
  Z.~Lin, N.~Gimelshein, L.~Antiga, A.~Desmaison, A.~Kopf, E.~Yang, Z.~DeVito,
  M.~Raison, A.~Tejani, S.~Chilamkurthy, B.~Steiner, L.~Fang, J.~Bai, and
  S.~Chintala, ``Pytorch: An imperative style, high-performance deep learning
  library,'' in \emph{Advances in Neural Information Processing Systems 32},
  2019, pp. 8024--8035.

\bibitem{cao2014crema}
H.~Cao, D.~Cooper, M.~Keutmann, R.~Gur, A.~Nenkova, and R.~Verma, ``Crema-d:
  Crowd-sourced emotional multimodal actors dataset,'' \emph{IEEE transactions
  on affective computing}, vol.~5, no.~4, pp. 377--390, 2014.

\bibitem{livingstone2018ryerson}
S.~Livingstone and F.~Russo, ``The ryerson audio-visual database of emotional
  speech and song (ravdess): A dynamic, multimodal set of facial and vocal
  expressions in north american english,'' \emph{PloS one}, vol.~13, no.~5, p.
  e0196391, 2018.

\bibitem{ringeval2013introducing}
F.~Ringeval, A.~Sonderegger, J.~Sauer, and D.~Lalanne, ``Introducing the recola
  multimodal corpus of remote collaborative and affective interactions,'' in
  \emph{2013 10th IEEE international conference and workshops on automatic face
  and gesture recognition (FG)}.\hskip 1em plus 0.5em minus 0.4em\relax IEEE,
  2013, pp. 1--8.

\bibitem{kossaifi2019sewa}
J.~Kossaifi, R.~Walecki, Y.~Panagakis, J.~Shen, M.~Schmitt, F.~Ringeval,
  J.~Han, V.~Pandit, A.~Toisoul, B.~W. Schuller \emph{et~al.}, ``Sewa db: A
  rich database for audio-visual emotion and sentiment research in the wild,''
  \emph{IEEE Transactions on Pattern Analysis and Machine Intelligence}, 2019.

\bibitem{busso2008iemocap}
C.~Busso, M.~Bulut, C.-C. Lee, A.~Kazemzadeh, E.~Mower, S.~Kim, J.~N. Chang,
  S.~Lee, and S.~S. Narayanan, ``Iemocap: Interactive emotional dyadic motion
  capture database,'' \emph{Language resources and evaluation}, vol.~42, no.~4,
  p. 335, 2008.

\bibitem{majumder2018multimodal}
N.~Majumder, D.~Hazarika, A.~Gelbukh, E.~Cambria, and S.~Poria, ``Multimodal
  sentiment analysis using hierarchical fusion with context modeling,''
  \emph{Knowledge-Based Systems}, vol. 161, pp. 124--133, 2018.

\bibitem{warden2018speech}
P.~Warden, ``Speech commands: A dataset for limited-vocabulary speech
  recognition,'' \emph{arXiv preprint arXiv:1804.03209}, 2018.

\bibitem{arandjelovic2018objects}
R.~Arandjelovic and A.~Zisserman, ``Objects that sound,'' in \emph{ECCV}, 2018,
  pp. 435--451.

\bibitem{korbar2018cooperative}
B.~Korbar, D.~Tran, and L.~Torresani, ``Cooperative learning of audio and video
  models from self-supervised synchronization,'' in \emph{NeurIPS}, 2018, pp.
  7763--7774.

\bibitem{eyben2013recent}
F.~Eyben, F.~Weninger, F.~Gross, and B.~Schuller, ``Recent developments in
  opensmile, the munich open-source multimedia feature extractor,'' in
  \emph{Proceedings of the 21st ACM international conference on Multimedia},
  2013, pp. 835--838.

\bibitem{ravanelli2018light}
M.~Ravanelli, P.~Brakel, M.~Omologo, and Y.~Bengio, ``Light gated recurrent
  units for speech recognition,'' \emph{IEEE Transactions on Emerging Topics in
  Computational Intelligence}, vol.~2, no.~2, pp. 92--102, 2018.

\bibitem{varga1993assessment}
A.~Varga and H.~J. Steeneken, ``Assessment for automatic speech recognition:
  Ii. noisex-92: A database and an experiment to study the effect of additive
  noise on speech recognition systems,'' \emph{Speech communication}, vol.~12,
  no.~3, pp. 247--251, 1993.

\bibitem{ephrat2018looking}
A.~Ephrat, I.~Mosseri, O.~Lang, T.~Dekel, K.~Wilson, A.~Hassidim, W.~Freeman,
  and M.~Rubinstein, ``Looking to listen at the cocktail party: A
  speaker-independent audio-visual model for speech separation,''
  \emph{SIGGRAPH}, 2018.

\bibitem{ravanelli2018interpretable}
M.~Ravanelli and Y.~Bengio, ``Interpretable convolutional filters with
  sincnet,'' \emph{IEEE SLT Workshop}, 2018.

\bibitem{vougioukas2019video}
K.~Vougioukas, P.~Ma, S.~Petridis, and M.~Pantic, ``Video-driven speech
  reconstruction using generative adversarial networks,'' \emph{Proc.
  Interspeech 2019}, pp. 4125--4129, 2019.

\bibitem{ma2020learning}
S.~Ma, Z.~Zeng, D.~McDuff, and Y.~Song, ``Learning audio-visual representations
  with active contrastive coding,'' \emph{arXiv preprint arXiv:2009.09805},
  2020.

\end{thebibliography}
% }
% \vfill\null

% \columnbreak


\begin{IEEEbiography}[{\includegraphics[width=1in,height=1.25in,clip,keepaspectratio]{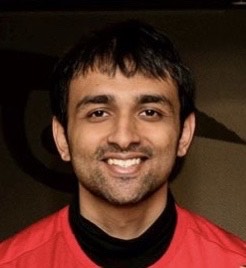}}]{Abhinav Shukla}
is a PhD student in the iBUG group at Imperial College London. He completed his Masters in Computer Science at IIIT Hyderabad, India, from where he also received a Bachelors degree in Computer Science and Engineering in 2017. He was also a visiting researcher at the SeSaMe Center at the National University of Singapore in 2018. He was an intern in the audio team at Facebook Reality Labs in 2020. His research interests include self-supervised and multimodal representation learning with applications in affective computing and audiovisual processing.
\end{IEEEbiography}

\begin{IEEEbiography}[{\includegraphics[width=1in,height=1.25in,clip,keepaspectratio]{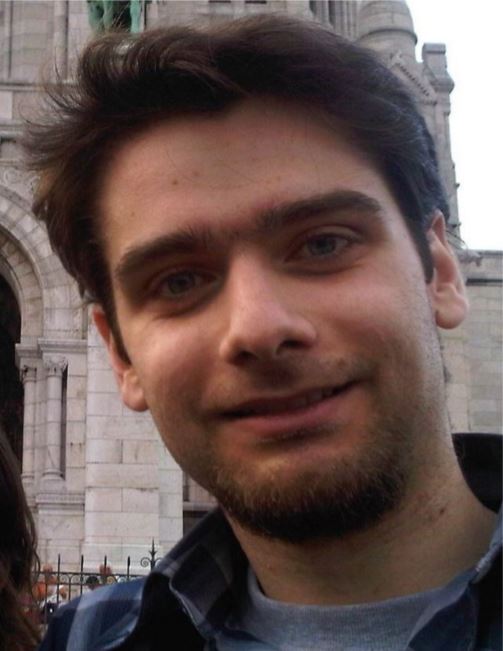}}]{Stavros Petridis}
is a research scientist at Facebook AI Applied Research and honorary research fellow at the Department of Computing at Imperial College London. He received his BSc degree in electrical and computer engineering from the Aristotle University of Thessaloniki, Greece in 2004 and his MSc degree in Advanced Computing and Ph.D. from Imperial College London in 2005 and 2012, respectively. He has been a research intern in the Image Processing Group at University College London and the Field Robotics Centre, Robotics Institute, Carnegie Mellon University and a visiting researcher at the affect analysis group at University of Pittsburgh. His research interests lie in the areas of pattern recognition and machine learning and their application to multimodal recognition of human non-verbal behaviour and non-linguistic vocalisations. He is currently working on deep learning approaches for audiovisual fusion. He is a member of the IEEE
\end{IEEEbiography}

\begin{IEEEbiography}[{\includegraphics[width=1in,height=1.25in,clip,keepaspectratio]{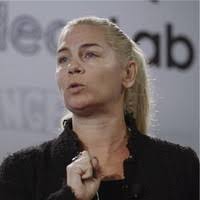}}]{Maja Pantic}
is a professor in affective and behavioral computing in the Department of Computing at Imperial College London, UK. She was the Research Director of Samsung AI Centre, Cambridge, UK from 2018 to 2020 and is currently an AI Scientific Research Lead at Facebook London. She currently serves as an associate editor for both the IEEE Transactions on  Pattern  Analysis and Machine Intelligence and the IEEE Transactions on Affective  Computing. She has received various awards for her work on automatic analysis of human behavior,including the Roger Needham Award 2011. She is a fellow of the UK’s Royal Academy of Engineering, the IEEE, and the IAPR
\end{IEEEbiography}


%\vfill


\end{document}